\documentclass[aps,showpacs,onecolumn,notitlepage,floatfix,amsmath,amssymb,nofootinbib]{revtex4-1}

\usepackage{bigints}
\usepackage{amssymb}
\usepackage{amsmath}
\usepackage{epsfig}
\usepackage{graphicx}
\usepackage[dvipsnames]{xcolor}
\usepackage{latexsym}
\usepackage{bm,latexsym}
\usepackage[mathscr]{eucal}
\usepackage{float} 
\usepackage{pbox}
\usepackage{ulem}

\begin{document}

\title{Spiral galaxy spacetime in extended scalar-tensor-Gauss-Bonnet gravity coupled to nonlinear electrodynamics}

\author{Pedro Ca\~nate} 
\email[]{pcannate@gmail.com }
\affiliation{Departamento de Ciencias Naturales, Facultad de Ciencias Exactas y Naturales, Universidad Surcolombiana, Avenida Pastrana Borrero - Carrera 1, A.A. 385, C.P. 410001, Neiva, Huila, Colombia }

\begin{abstract} 
In this work, using nonlinear electrodynamics (NLED) as a source in the field equations of extended scalar-tensor-Gauss-Bonnet (E-STGB) gravity, a novel type of static and spherically symmetric spacetime is found. 
We show that this novel type of spacetime can be used to explain the rotation curves and gravitational lensing of spiral galaxies.
The NLED model used to derive our solution, in the weak and strong field limits, reduces to Maxwell electrodynamics. Furthermore, for a wide range of parameter values where our solution describes the gravitational field of a spiral galaxy, the NLED model satisfies the null and weak energy conditions.
Hence, our results suggest that the extended scalar-tensor-Gauss-Bonnet gravity minimally coupled to nonlinear electrodynamics (E-STGB-NLED) provides a viable alternative to explain flat rotation curves of spiral galaxies without invoking dark matter.
\end{abstract}
\pacs{04.20.Jb, 04.50.Kd, 04.50.-h, 04.60.Cf }
%

\maketitle

\section{Introduction} 
Over ninety years ago, the dark matter (DM) paradigm ~--a hypothetical form of matter that interacts very weakly 
with ordinary matter and electromagnetic radiation, but yet has significant gravitational effects on them--~
was originally proposed by Zwicky in his landmark work \cite{Zwicky}, in which,  by observing the excessive rotational velocity of luminous matter in the Coma galaxy cluster compared to the calculated mass density of the cluster, he proposed that the circular motion of the neutral hydrogen gas clouds around the center of most galaxies cannot be explained in terms of the luminous matter content of the galaxies, at least not within the framework of  
standard gravitational theory.  
Later in the 1970s, by Rubin \cite{Rubin79a,Rubin79b,Rubin83}, it was verified that the observed rotation curves\footnote{ A disk galaxy (or disc galaxy) is a kind of galaxy that has a galactic disk, which consists of a flattened circular volume of stars and hydrogen gas clouds  that orbit in a circular coplanar motion around the galactic center. 
The spiral galaxies are a prominent subclass of disk galaxies,  distinguished by their signature spiral arms extending from their galactic center. The rotation curve of a disk galaxy (also known as velocity curve) is a plot of the tangential velocities (or orbital speeds) of visible stars or hydrogen gas clouds in the galaxy versus their radial distance from the center of the galaxy. For a review of spiral galaxies and their features, see Ref. \cite{Binney1987,Bhattacharjee2014}.} for spiral galaxies did not follow the expected Keplerian behavior in the outer regions of galaxies. 
In these galaxies, the observed ``tangential velocity''  $v(r)$  of stars and hydrogen gas clouds in circular orbits  (of radius $r$) around the galactic center up to distances far beyond the luminous radius of these galaxies, is found to be approximately constant (flat) or slightly rising with increasing $r$, defining a region  known as the flat rotation curves (FRC) zone\footnote{In other words, in the FRC zone of a spiral galaxy, the tangential velocity $v(r)$ of stars and hydrogen gas clouds on a circular orbit of radius $r$ around the galactic center is observed to be approximately constant (or slightly rising) and independent of their radial distance, {\it i.e.,} $v(r)\approx constant$.}. 
More specifically, according to the currently most accepted theory of gravity on galactic scales,
the tangential velocity of a non-relativistic massive (test) particle in a circular orbit (of radius $r$) around the galactic center of a spiral galaxy can be expressed as $v(r) =\sqrt{GM(r)/r}$, where $M(r)$ represents the total (galactic) mass inside the orbit.
This indicates that the prevailing theory is consistent with the observed rotation curves, provided that it is assumed $M(r)$ increases linearly with $r$ (even at large distances, where very little luminous matter has been detected) suggesting  the existence of certain type of non-luminous (invisible) matter (commonly referred to as gravitational dark matter) distributed in a spherical halo around spiral galaxies. 
That is, $M(r)=v^{2}_{0}r/G$, which implies a (galactic) mass-energy density given by $\rho(r) = v^{2}_{0}/\!(4\pi G r^{2})$ being $v_{0}$ a constant that, depending on the spiral galaxy, can have a value between 200 and 300 $km/s$ (see Ref. \cite{Ostriker74} for a review). \\
On the other hand, regarding the precise nature of dark matter, within the context of particle physics that extends beyond the Standard Model (BSM) numerous candidates for dark matter have been proposed, being the Weakly Interacting Massive Particles (WIMPs) and the axions considered the most popular. 
Other dark matter candidates in BSM are the Inert Higgs Doublet, sterile neutrinos and Kaluza-Klein particles  
(see \cite{Bergstrom09} for a review). Multiple observational and experimental approaches have been utilized in an attempt to identify dark matter particles \cite{Drukier86}. 
However, despite extensive efforts, the detection of dark matter remains a challenging issue due to its extremely weak interactions with conventional matter and light, creating doubts about its existence.
But, if DM hypothesis is excluded, the existence of flat rotation curves in spiral galaxies suggests a potential discrepancy with predictions from Einstein’s (and the Newtonian) gravity on galactic and larger scales.
Based on this premise, M. Milgrom proposed Modified Newtonian Dynamics (MOND) \cite{Milgrom83}, which basically is a modification of Newton's laws of motion and gravitation to explain the flat rotation curves of spiral galaxies without invoking dark matter\footnote{{\it I.e.,} the MOND gravity provides a straightforward explanation (without the dark matter paradigm) for why the tangential velocity of test particles (such as stars and hydrogen clouds) orbiting the galactic center is approximately constant and independent of their radial distance. }. 
Since MOND gravity by itself is not a theory, numerous efforts have been undertaken to develop a theory for MOND,  
either relativistic or non-relativistic (see Refs. \cite{Bekenstein84,Bekenstein05} for a review).
Nevertheless, the MOND paradigm has not yet been transformed into a fully relativistically invariant theory. 
Additionally, numerous extensions to general relativity (GR), also known as modified theories of gravity, have been proposed as potential ways to circumvent the need for dark matter.  These include theories with higher derivative gravity (such as $f(R)$ gravity \cite{Capozziello06,Harko10}) and theories with non-minimally coupled scalar fields to gravity (see for instance \cite{Nucamendi2001}). 

In this work, we address this issue using a novel type of asymptotically flat (AF) static and spherically symmetric (SSS) purely magnetic solution of the E-STGB theory minimally coupled to NLED.
The E-STGB theory is based on the established $D$-dimensional Einstein-Gauss-Bonnet (EGB) gravity, with the addition of a scalar field non-minimally coupled to the Gauss-Bonnet invariant. 
EGB gravity represents a fascinating modified 
gravity theory, proving to be one of the most successful extensions of higher-dimensional Einstein gravity in the sector of higher-curvature \cite{Lovelock,Lanczos38}. \\
The NLED theories have attracted interest since the beginning of the  
{\it 20th} century, mainly because it may be able to solve problems that traditional Maxwell electrodynamics is unable to explain (for a recent review, see \cite{Sorokin2022}). 
More precisely,  the NLED  theories are derived from Lagrangians $\mathcal{L}=\mathcal{L}(\mathcal{F}, \mathcal{G})$ that depend arbitrarily on the two Lorentz invariants $\mathcal{F}$ and $\mathcal{G}$. These invariants are quadratic in the electric  $\boldsymbol{\mathcal{E}}$ and magnetic $\boldsymbol{\mathcal{B}}$ fields and are generally defined as: $\mathcal{F}= 2(\boldsymbol{\mathcal{B}}^2-\boldsymbol{\mathcal{E}}^2)$ and $\mathcal{G} = \boldsymbol{\mathcal{E}}\cdot\boldsymbol{\mathcal{B}}$. 
The NLED theory proposed by Born and Infeld (BI) in 1933, which sought to remove the singularities related to the self-energy of point charges, was the first notable model \cite{Born1933}. The Lagrangian that describes the BI theory is given by 
\begin{equation}\label{BI_FG}
\mathcal{L}_{_{\mathrm{BI}}}\!(\mathcal{F},\mathcal{G}) = 4b^{2} \left( -1 + \sqrt{ 1 + \frac{\mathcal{F}}{2b^{2}} + \frac{\mathcal{G}^{2}}{16b^{4}} }\!~\right)
\end{equation}
where  $b$ is a constant which has the physical interpretation of a critical field
strength. \\
A few years later, in 1935, Euler and Heisenberg (EH) developed a novel NLED model which computed a complete effective action describing nonlinear corrections to Maxwell's theory due to quantum electron-positron one-loop effects \cite{Heuler_35,Heisenberg_36}. In a nonperturbative form, the Lagrangian that describes the EH theory is expressed as 
\begin{equation}\label{EH}
\mathcal{L}_{_{\mathrm{EH}}}  
= \mathcal{L}_{_{_{\mathrm{LED}}}} + \frac{1}{8\pi^{2}}\int^{^{\infty}}_{_{0}} \frac{ e^{-m^{2}_{e}\varrho} }{\varrho^{3}}\left[ \left(q_{e}\varrho\right)^{2}\frac{ {\tt Re}\!\left\{ \cosh\left( q_{e}\varrho \sqrt{ 2\mathcal{F}+ 2i\mathcal{G} }  \right) \right\} }{{\tt Im}\!\left\{ \cosh\left( q_{e}\varrho \sqrt{ 2\mathcal{F}+ 2i\mathcal{G} }  \right) \right\}}\mathcal{G} - \frac{2}{3}\left(q_{e}\varrho\right)^{2}\mathcal{F} - 1 \right] d\varrho
\end{equation}
where $m_{e}$ is the mass of the electron and $q_{e}$ is the elementary charge. \\
Nowadays, NLED theories are being investigated due to their capacity to explain high-intensity electromagnetic fields and their potential applications in a variety of physics fields, including cosmology and astrophysics  
(see, for instance, Refs. \cite{Garcia2000,Lorenci2002,Breton2012}). Furthermore, considering NLED Lagrangians of the form $\mathcal{L}(\mathcal{F})$ coupled to gravity, interesting solutions arise, like regular black holes and $(2+1)$-dimensional traversable wormholes, among others \cite{Ayon2000,Ayon_Garcia,Canate_Breton,Novello2004,Campanelli2008}. 

Throughout this paper, our conventions are: The geometric unit system, {\it i.e.,} the speed of light in vacuum $c$, and the gravitational constant $G_{0}$, are set equal to unity $G_{0}=c=1$. In addition, the signature of
the metric is $(-,+,+,+)$, and we shall consider NLED Lagrangians that only depend on the invariant $\mathcal{F}$.

\section{From D-dimensional Einstein-Gauss-Bonnet gravity to extended scalar-tensor-Gauss-Bonnet theories}
The action of $D$-dimensional EGB gravity with additional matter fields, is given by
\begin{equation}\label{actionGB}
S_{_{\rm EGB}}[g_{\mu\nu}, \boldsymbol{\psi}_{a}] = \frac{1}{16\pi}\int d^{^{D}}\!\!x \sqrt{-g} \left[ R + \alpha R_{_{G\!B}}^{2} + \mathcal{L}_{\rm matter}(g_{\mu\nu},\boldsymbol{\psi}_{a})\right]
\end{equation}
where $R$ is the Ricci scalar curvature, $\alpha$ is a coupling constant with dimension $[\textup{lengh}]^{2}$ known as the Gauss-Bonnet (GB) coefficient (also called second order Lovelock coefficient), $R_{_{G\!B}}^{2}$ stands for the quadratic GB invariant, which is a higher curvature term and is defined in terms of the curvature invariants as $R_{_{G\!B}}^{2} = R_{\sigma\beta\mu\nu}R^{\sigma\beta\mu\nu} - 4R_{\sigma\beta}R^{\sigma\beta} + R^{2}$, 
where $R_{\sigma\beta\mu\nu}$ and $R_{\sigma\beta}$ represent the Riemann and Ricci tensor components, respectively.
The last term in the action is the matter fields Lagrangian, $\mathcal{L}_{\rm matter}$, where $\boldsymbol{\psi}_{a}$ describes any matter field included in the theory. 
The EGB theory is gaining popularity due to the fact that the field equations for $D-$dimensional EGB are second order partial differential equations for the metric tensor, similar to GR. Additionally, the effects of $R_{_{G\!B}}^{2}$ have the potential to alleviate some of the negative aspects of GR, such as singularities, and for instance in \cite{Cosm_GB} it is demonstrated that in EGB the singularity in $(4+1)-$dimensional Friedmann-Robertson-Walker cosmology can be removed; while in \cite{Cai} and \cite{Anabalon} novel $D$-dimensional black hole solutions (with $D>4$), that have no  GR limit, were reported. \\
In contrast, EGB theory is precisely equivalent to general relativity for $(3+1)-$dimensional spacetimes, as per Lovelock's theorem \cite{Lovelock}.
However, within this frame, if a canonical massless scalar field $\phi$ is non-minimally coupled to the GB invariant, {\it i.e.,} promoting the GB coefficient $\alpha$ to a GB coupling function that depends only on  
$\phi$, that is $\alpha=\alpha(\phi)$,  one obtains the general action of extended scalar-tensor-Gauss-Bonnet (E-STGB) theories, also referred to as Einstein-scalar-Gauss-Bonnet gravity (EsGB), given by
\begin{equation}\label{actionESTGB}
S_{_{\rm ESTGB}}[g_{\mu\nu}, \phi,\boldsymbol{\psi}_{a}] = \int d^{^{4}}\!\!x \sqrt{-g} \left\{\frac{1}{16\pi}\left[R - \frac{1}{2}\nabla_{\mu}\phi\nabla^{\mu}\phi + \alpha(\phi) R_{_{G\!B}}^{2}\right] + \mathcal{L}_{\rm matter}(g_{\mu\nu},\boldsymbol{\psi}_{a}) \right\}.
\end{equation}
The field equations derived from this action continue being of second order partial differential equations for the metric tensor; however, they are significantly different from the conventional GR field equations.
Thus, E-STGB gravity emerges as one of the best four-dimensional modify-GR theories in the sector of higher-curvature (see \cite{scalariz_Doneva} for details).
Because of this, in high-curvature regions the predictions of E-STGB theories diverge significantly from those of GR, such as the interior of black holes and the early universe epochs, where E-STGB gravity aims to resolve curvature singularities. 
Examples include rotating black holes that generalize Kerr black holes \cite{Kunz2011}, and  
AF-SSS novel black holes with non-trivial scalar hair \cite{Kanti2018} that evade the no-hair theorems \cite{Bekenstein72,Sudarsky95}, all these numerically derived in context of E-STGB gravity.  
In these theories, spontaneous and induced scalarization of black holes arise \cite{scalariz_Doneva}; the stability of those scalarized black holes has been addressed in \cite{Kunz2018}.
Furthermore, E-STGB gravity has recently drawn more attention as a viable candidate 
to explain the current stage of cosmic acceleration \cite{cosmic_acceleration}.

Regarding exact solutions, using linear/nonlinear electromagnetic fields as sources in the E-STGB field equations has allowed for the creation of interesting exact solutions, such as AF-SSS traversable wormholes (without need of exotic matter) and novel AF-SSS black holes (see Refs. \cite{Canate19A,Canate19,Canate22,Canate23,Canate21}).
More exactly, in this modified gravity framework, it has been sufficient to consider either purely electric configurations with electromagnetic invariant $\mathcal{F} \sim -\mathcal{E}^{2}$ (where $\mathcal{E}$ is the electric field vector) \cite{Canate19A,Canate19}, or purely magnetic configurations with $\mathcal{F} \sim \mathcal{B}^{2}$ (where $\mathcal{B}$ is the magnetic field vector) \cite{Canate22,Canate23,Canate21}, in order to generate new exact AF-SSS charged solutions. 
Another type of exact solution has been derived in \cite{Canate21B}, which consists of a curve spacetime analogous to the canonical acoustic black hole, which arises when considering a self-interacting scalar field non-minimally
coupled to the Gauss-Bonnet invariant. 

\subsection{EXTENDED SCALAR-TENSOR-GAUSS-BONNET THEORY MINIMALLY COUPLED TO NONLINEAR ELECTRODYNAMICS}
The extended scalar-tensor-Gauss-Bonnet theory minimally coupled to
nonlinear electrodynamics (E-STGB-NLED) is defined by the following action
\begin{equation}\label{actionLED}
S_{_{\rm ESTGB}}[g_{\mu\nu},\phi,A_{_{\beta}}] = \int d^{4}x \sqrt{-g} \left\{ \frac{1}{16\pi}\left[R - \frac{1}{2}\nabla_{\mu}\phi\nabla^{\mu}\phi + \alpha(\phi) R_{_{GB}}^{2} \right] - \frac{1}{4\pi}\mathcal{L}(\mathcal{F}) \right\}
\end{equation}
which is basically the action \eqref{actionESTGB} with $\mathcal{L}_{\rm matter} = - \mathcal{L}(\mathcal{F})/(4\pi)$, being  $\mathcal{L}(\mathcal{F})$ a function of the electromagnetic invariant $\mathcal{F}\equiv \frac{1}{4}F_{\mu\nu}F^{\mu\nu}$, where $F_{\mu\nu}=2\partial_{[\mu}A_{\nu]}$ are the components of the electromagnetic field tensor $\boldsymbol{F}=\frac{1}{2}F_{\mu\nu} \boldsymbol{dx^{\mu}} \wedge \boldsymbol{dx^{\nu}}$, and $A_{\nu}$ the components of the electromagnetic potential $\boldsymbol{A}=A_{\nu}\boldsymbol{dx^{\nu}}$. 
For instance, the standard formulation of classical electrodynamics, often called  Maxwell electrodynamics, is based on a linear electromagnetic Lagrangian density $\mathcal{L}_{_{Max}}(\mathcal{F})=\mathcal{F}$; while the electromagnetic Lagrangian density of Born-Infeld theory (BI), depending only on $\mathcal{F}$, 
is given by
\begin{equation}\label{BI}
\mathcal{L}_{_{\mathrm{BI}}}\!(\mathcal{F}) = 4b^{2} \left( -1 + \sqrt{ 1 + \frac{\mathcal{F}}{2b^{2}} }\right)
\end{equation} 
which  to keep the NLED Lagrangian real (positive term under the square root), the invariant $\mathcal{F}$ must satisfy $\mathcal{F}\geq - 2b^{2}$.
Varying the action \eqref{actionLED} with respect to the metric tensor $g^{\mu\nu}$,  
we get the following field equations  
\begin{equation}\label{EQ_gaa}
G_{\mu}{}^{\nu} =  8\pi(T_{\mu}{}^{\nu})\!_{_{_{S\!G\!B}}} +  8\pi(T_{\mu}{}^{\nu})\!_{_{_{N\!L\!E\!D}}}, %
\end{equation}
where $G_{\mu}{}^{\nu}=R_{\mu}{}^{\nu} - \frac{R}{2}\delta_{\mu}{}^{\nu}$ denotes the components of the Einstein tensor, and the quantities $(T_{\mu}{}^{\nu})\!_{_{_{S\!G\!B}}}$ and $(T_{\mu}{}^{\nu})\!_{_{_{N\!L\!E\!D}}}$ are defined by the following expressions:
\begin{eqnarray}
&&8\pi (T_{\mu}{}^{\nu})\!_{_{_{S\!G\!B}}} =  \frac{1}{2}\nabla_{\mu}\phi \nabla^{\nu}\phi - \frac{1}{4}(\nabla_{\beta}\phi \nabla^{\beta}\phi)\delta_{\mu}{}^{\nu}  
- \frac{1}{2}( g_{\mu\rho} \delta_{\lambda}{}^{\nu} + g_{\mu\lambda} \delta_{\rho}{}^{\nu})\eta^{\chi\lambda\gamma\sigma}\tilde{R}^{\rho\beta}{}_{\gamma\sigma}\nabla_{\beta}\nabla_{\chi}\alpha(\phi) \\
&& 8\pi(T_{\mu}{}^{\nu})\!_{_{_{N\!L\!E\!D}}}  
= 2\mathcal{L}_{_{\mathcal{F}}}F_{\mu\beta}F^{\nu\beta} - 2\mathcal{L}\hskip.06cm\delta_{\mu}{}^{\nu} \label{NLED_EM}
\end{eqnarray}
with $\tilde{R}^{\rho\beta}{}_{\gamma\sigma} = \eta^{\rho\beta\lambda\tau}R_{\lambda\tau\gamma\sigma} = \epsilon^{\rho\beta\lambda\tau}R_{\lambda\tau\gamma\sigma}/\sqrt{-g}$, while $\nabla_{\chi}$ is the covariant derivative with respect to the spacetime metric $g_{\mu\nu}$, and $\mathcal{L}_{_{\mathcal{F}}} = \frac{d\mathcal{L}}{d\mathcal{F}}$. 
The quantities $(T_{\mu}{}^{\nu})\!_{_{_{S\!G\!B}}}$ denote the components of the Scalar-Gauss-Bonnet (SGB) tensor, which describes the effects of the scalar field $\phi$ and the higher-curvature term $\alpha(\phi) R_{_{GB}}^{2}$ on spacetime geometry.
These components can be expressed as
\begin{equation}\label{SGB_EM}
8\pi (T_{\mu}{}^{\nu})\!_{_{_{S\!G\!B}}} = 8\pi (T_{\mu}{}^{\nu})\!_{_{_{S\!F}}} + \Gamma_{\mu}{}^{\nu}    
\end{equation}
where the components $(T_{\mu}{}^{\nu})\!_{_{_{S\!F}}}$ and $\Gamma_{\mu}{}^{\nu}$ are given by
\begin{equation}\label{SGB_ten}
8\pi (T_{\mu}{}^{\nu})\!_{_{_{S\!F}}} = \frac{1}{2}\nabla_{\mu}\phi \nabla^{\nu}\phi - \frac{1}{4}(\nabla_{\beta}\phi \nabla^{\beta}\phi)\delta_{\mu}{}^{\nu}, \quad 
\quad
\Gamma_{\mu}{}^{\nu} = - \frac{1}{2}( g_{\mu\rho} \delta_{\lambda}{}^{\nu} + g_{\mu\lambda} \delta_{\rho}{}^{\nu})\eta^{\chi\lambda\gamma\sigma}\tilde{R}^{\rho\beta}{}_{\gamma\sigma}\nabla_{\beta}\nabla_{\chi}\alpha(\phi). 
\end{equation}
This indicates that the tensor
$(T_{\mu}{}^{\nu})\!_{_{_{S\!G\!B}}}$ consists of a term
$(T_{\mu}{}^{\nu})\!_{_{_{S\!F}}}$ that depends only on the scalar field $\phi$, in addition to a second term 
$\Gamma_{\mu}{}^{\nu}$ related to the coupling between $\phi$ and the GB curvature.
Therefore, when $\alpha(\phi)$ represents a constant function, yields $\Gamma_{\mu}{}^{\nu}=0$, 
resulting in the tensor $(T_{\mu}{}^{\nu})\!_{_{_{S\!G\!B}}}$ becoming the standard energy-momentum tensor associated with a massless scalar field, referred to as $(T_{\mu}{}^{\nu})\!_{_{_{S\!F}}}$.
\\
The components of the NLED energy-momentum tensor are denoted by $(T_{\mu}{}^{\nu})\!_{_{_{N\!L\!E\!D}}}$.
On the other hand, the equations of motion for the nonlinear electrodynamics theory are as follows
\begin{equation}\label{NLED_Eq}
\nabla_{\mu}(\mathcal{L}_{_{\mathcal{F}}}F^{\mu\nu})=0,  \quad\quad  d\boldsymbol{F} = 0.
\end{equation}
By taking the divergence of Eq. \eqref{EQ_gaa}, taking into account that $\nabla_{\nu}(T_{\mu}{}^{\nu})\!_{_{_{N\!L\!E\!D}}} =0$, and the knowledge that $\nabla_{\nu}G_{\mu}{}^{\nu} =0$ is guaranteed by Bianchi identities, it follows that $\nabla_{\nu}(T_{\mu}{}^{\nu})\!_{_{_{S\!G\!B}}} =0$. This in turn implies that 
\begin{equation}\label{scalar_Eq}
\nabla^{2}\phi + \dot{\alpha} R_{_{G\!B}}^{2}  = 0,
\end{equation}
where $\dot{\alpha}:= \frac{d\alpha}{d\phi}$ (similar notation will be used for higher derivatives).  
Thus, the motion equations \eqref{EQ_gaa}, \eqref{NLED_Eq} and \eqref{scalar_Eq} constituting the fields equations of E-STGB-NLED theory.
For the particular case $\alpha(\phi)=const$ ({\it i.e.,} reduction to 4D-EGB theory minimally coupled to a massless scalar field) yields $\dot{\alpha}=0=\Gamma_{\mu}{}^{\nu}$ and then Eq. \eqref{EQ_gaa} becomes $G_{\mu}{}^{\nu} = 8\pi (T_{\mu}{}^{\nu})\!_{_{_{S\!F}}} + 8\pi (T_{\mu}{}^{\nu})\!_{_{_{N\!L\!E\!D}}}$, Eq. \eqref{NLED_Eq} is not affected, while Eq. \eqref{scalar_Eq} takes the form $\nabla^{2}\phi=0$, indicating that in this case 
the E-STGB-NLED theory reduces to Einstein-nonlinear electrodynamics gravity coupled to a massless scalar field (GR-NLED-SF). 

\subsection{E-STGB-NLED FIELD EQUATIONS FOR PURELY MAGNETIC SSS SPACETIMES}
The most general metric for a SSS spacetime solution of E-STGB-NLED, can be expressed in terms of the line element
\begin{equation}\label{SSSmet}
ds^{2} =  - e^{ A(r) }dt^{2} + e^{ B(r) }dr^{2}  + r^{2}d\Omega^{2}    
\end{equation}
with $d\Omega^{2} = d\theta^{2}  + \sin^{2}\theta d\varphi^{2}$,   
where $A = A(r)$ and $B = B(r)$ are functions to be determined from the E-STGB-NLED 
field equations. For this general SSS configuration, the nonvanishing components of the Einstein tensor can be expressed as
\begin{equation}\label{GabSSS}
G_{t}{}^{t}\!=\!\frac{ e^{^{\!\!-B}} }{ r^{2} }\!\!\left( -rB' - e^{^{\!B}} + 1 \right)\!, \hspace{0.25cm}  G_{r}{}^{r} \!=\! \frac{ e^{^{\!\! -B}} }{ r^{2} }\!\!\left( rA' - e^{^{\!B}} + 1 \right)\!, \hspace{0.25cm} G_{\theta}{}^{\theta}\!=\! G_{\varphi}{}^{\varphi}\!=\!\frac{ e^{^{\!\!-B}} }{ 4r }\!\!\left( rA'^{2} - rA'B' + 2rA'' + 2A' - 2B' \right),
\end{equation}
where  $A':=\frac{dA}{dr}$, we shall use a similar notation for higher derivatives. 
On the other hand, the nontrivial components of $(T_{\mu}{}^{\nu})\!_{_{_{S\!G\!B}}}$ are
\begin{eqnarray}
&& 8\pi(T_{t}{}^{t})\!_{_{_{S\!G\!B}}} = -\frac{ e^{ -2B} }{ 4r^{2} }\left\{ r^{2}e^{B}\phi'^{2} + 16(e^{B} - 1)\phi'^{2} \ddot{\alpha} - 8[ (e^{B} - 3)B'\phi' - 2(e^{B} - 1)\phi'']\dot{\alpha} \right\}, \label{pGBdttt}\\
&& 8\pi(T_{r}{}^{r})\!_{_{_{S\!G\! B}}} = \frac{ e^{ -B} }{ 4 } \left[\phi'^{2}  - \frac{8(e^{B} - 3)e^{-B}A' \phi'\dot{\alpha} }{r^{2}}\right],  \label{pGBdrrr} \\
&& 8\pi(T_{\theta}{}^{\theta})\!_{_{_{S\!G\! B}}} = 8\pi(T_{\varphi}{}^{\varphi})\!_{_{_{S\!G\! B}}} = -\frac{ e^{ -2B} }{ 4r } \left\{ re^{B}\phi'^{2} - 8A'\phi'^{2}\ddot{\alpha} - 4\left[ (A'^{2} + 2A'')\phi' + (2\phi'' - 3B'\phi')A'\right]\dot{\alpha} \right\}. \label{pGBdrthth}
\end{eqnarray}
where, due to static and spherical symmetry, the scalar field becomes function only of the radial coordinate, that is $\phi=\phi(r)$. Consequently, $\alpha=\alpha(\phi)$ can be written as a function that depends only on $r$,
{\it i.e.,} $\alpha=(\alpha\circ\phi)(r)$.   
Therefore, as long as $\phi'\neq0$ (non-trivial scalar field), follow that $\dot{\alpha}=\frac{\alpha'}{\phi'}$ and $\ddot{\alpha} = \frac{\alpha''}{\phi'^{2}} - \frac{\alpha'\phi''}{\phi'^{3}}.$  
Thus,  in the context of static and spherically symmetric spacetimes, the components \eqref{pGBdttt}$-$\eqref{pGBdrthth} can be written as
\begin{eqnarray} 
&&8\pi(T_{t}{}^{t})\!_{_{_{S\!G\!B}}}  
= - \frac{1}{4}e^{ -B}\phi'^{2}  - \frac{ e^{ -2B} }{ r^{2} }\left\{  4(e^{B} - 1)\alpha'' - 2(e^{B} - 3)B'\alpha' \right\},  \label{GBdttt} \\ 
&&8\pi(T_{r}{}^{r})\!_{_{_{S\!G\!B}}}  
=  \frac{1}{4}e^{ -B} \phi'^{2}  - \frac{2(e^{B} - 3)e^{-2B}A' \alpha' }{r^{2}},   
\label{GBdrrr} \\
&&8\pi(T_{\theta}{}^{\theta})\!_{_{_{S\!G\!B}}}  
=8\pi(T_{\varphi}{}^{\varphi})\!_{_{_{S\!G\!B}}}  
= - \frac{1}{4}e^{ -B} \phi'^{2}  + \frac{ e^{ -2B} }{ r } \left[   2A'\alpha'' + \left(A'^{2} + 2A'' - 3B'A'\right)\alpha' \right]. 
\label{GBdrthth}
\end{eqnarray}
Regarding the electromagnetic field tensor, since the spacetime is static and spherically symmetric, then the only nonvanishing terms are the electric component $F_{tr}=\mathcal{E}(r)$ and the magnetic component $F_{\theta\varphi}=\mathcal{B}(r,\theta)$. However, in this work, we can restrict ourselves to the purely magnetic field, {\it i.e.,} $F_{tr} = 0$ and $F_{\theta\varphi} \neq 0$.
With this restriction, the components of the electromagnetic field tensor can be expressed as  $F_{\alpha\beta} = \left( \delta^{\theta}_{\alpha}\delta^{\varphi}_{\beta} - \delta^{\varphi}_{\alpha}\delta^{\theta}_{\beta} \right)F_{\theta\varphi}$.  
In this way, for a static and spherically symmetric spacetime with line element \eqref{SSSmet}, the general solution of the equations $\nabla_{\mu}(\mathcal{L}_{_{\mathcal{F}}}F^{\mu\nu})=0$ is
\begin{equation}\label{fabSOL}
F_{\theta\varphi} =  r^{4} \mathrm{Q}(r) \sin\theta,
\end{equation}
where $\mathrm{Q}(r)$ is a function of $r$ only. 
This means $\boldsymbol{F} = r^{4} \mathrm{Q}(r)\sin\theta \hspace{0.1cm} \boldsymbol{d\theta} \wedge \boldsymbol{d\varphi}$. Therefore, according to $d\boldsymbol{F} = 0$, 
\begin{equation}
d\boldsymbol{F} = \left(r^{4} \mathrm{Q}(r)\right)' \sin\theta \hspace{0.1cm} \boldsymbol{dr} \wedge \boldsymbol{d\theta} \wedge \boldsymbol{d\varphi} = 0    
\end{equation}
yields $\mathrm{Q}(r)= q/r^{4}$, where $q$ is an integration constant associated with  
magnetic charge. 
Hence, for purely magnetic configurations,  the components of the electromagnetic field tensor $F_{\alpha \beta}$ and the invariant $\mathcal{F}$ are given by
\begin{equation}\label{magnetica}
F_{\alpha\beta} =   q  \sin \theta \hspace{0.05cm} \left( \delta^{\theta}_{\alpha}\delta^{\varphi}_{\beta} - \delta^{\varphi}_{\alpha}\delta^{\theta}_{\beta} \right),  \quad\quad\quad\quad \mathcal{F} =  \frac{q^{2}}{2r^{4}}.
\end{equation}
Finally, the nonvanishing components of $(T_{\alpha}{}^{\beta})\!_{_{_{N\!L\!E\!D}}}$ for an arbitrary $\mathcal{L}(\mathcal{F})$ function, considering the standard static and spherically symmetric line element ansatz \eqref{SSSmet} and a purely magnetic field \eqref{magnetica}, are written as
\begin{equation}\label{E_nled}
8\pi (T_{t}{}^{t})\!_{_{_{N\!L\!E\!D}}} = 8\pi (T_{r}{}^{r})\!_{_{_{N\!L\!E\!D}}} =  -2\mathcal{L},   
\quad\quad\quad 8\pi (T_{\theta}{}^{\theta})\!_{_{_{N\!L\!E\!D}}} = 8\pi (T_{\varphi}{}^{\varphi})\!_{_{_{N\!L\!E\!D}}} =  2(2\mathcal{F}\mathcal{L}_{_{\mathcal{F}}} - \mathcal{L}). 
\end{equation}
Inserting the above given components into the field equations \eqref{EQ_gaa} written as $\mathrm{C}_{\mu}{}^{\nu} = G_{\mu}{}^{\nu} - 8\pi(T_{\mu}{}^{\nu})\!_{_{_{S\!G\!B}}} - 8\pi (T_{\mu}{}^{\nu})\!_{_{_{N\!L\!E\!D}}} =0$, we see that the  E-STGB-NLED field equations for SSS purely magnetic spacetimes\footnote{Static and spherically symmetric spacetime solutions of E-STGB-NLED theory in which the electromagnetic field tensor has only magnetic part.}, take the form:   
\begin{eqnarray}   
&&\mathrm{C}_{t}{}^{t} \!=\!0 \quad\Rightarrow\quad 4e^{^{\!B}}\left( -rB' - e^{^{\!B}} + 1 \right)  + 8\left[2(e^{^{\!B}} - 1)\alpha'' - (e^{^{\!B}} - 3)B'\alpha' \right] + r^{2} e^{^{\!B}} \phi'^{2} +  8r^{2} e^{^{\!2B}}\mathcal{L}
= 0,
\label{LFeqt}\\ 
&&\mathrm{C}_{r}{}^{r}\!=\!0 \quad\Rightarrow\quad 4e^{^{\!B}} \left( rA' - e^{^{\!B}} + 1 \right)  +  8(e^{^{\!B}} - 3)A'\alpha'  -  r^{2} e^{^{\!B}} \phi'^{2} +  8r^{2} e^{^{\!2B}}\mathcal{L}  
= 0,
\label{LFeqr}   
\end{eqnarray}
and the equations $\mathrm{C}_{\ell}{}^{\ell}=0$, with $\ell= \{\theta, ~\varphi\}$, both reduce to   
\begin{equation}\label{LFeqphi}
e^{^{\!B}} \left( rA'^{2} - rA'B' + 2rA'' + 2A' - 2B' \right)  - 4\left[   2A'\alpha'' + \left(A'^{2} + 2A'' - 3B'A'\right)\alpha' \right] +  r e^{^{\!B}} \phi'^{2} - 8re^{^{\!2B}}(2\mathcal{F}\mathcal{L}_{_{\mathcal{F}}} - \mathcal{L}) = 0. 
\end{equation}
Whereas, for this configuration, multiplying the equation \eqref{scalar_Eq} by $\phi'\neq0$ gives
\begin{equation}\label{GB_eq}
r(\phi'^{2})' + (4 + rA' - rB')\phi'^{2} + \frac{e^{-B}}{r}\left[ (e^{B} - 3)A'B' - (e^{B} - 1)(2A'' + A'^{2})\right] \alpha' = 0.
\end{equation}  
Note that the Eqs. \eqref{LFeqt}$-$\eqref{GB_eq} [with $\alpha= f(\phi)$, $\alpha' =  \phi'\dot{f}$ and $\alpha'' = \phi''\dot{f} + \phi'^{2}\ddot{f}\!~$] are the same equations (30)$-$(33) that were utilized in Ref. \cite{Canate23} to introduce a simple method to generate magnetically charged ultrastatic traversable wormholes without exotic matter.\\
On the other hand, in the context of general relativity minimally coupled to nonlinear electrodynamics (GR-NLED), purely magnetic\footnote{{\it I.e.,} with electromagnetic field tensor given by $\boldsymbol{F} = q\sin\theta \hspace{0.1cm} \boldsymbol{d\theta} \wedge \boldsymbol{d\varphi}$ being $q$ the magnetic charge.} AF-SSS spacetimes (that is,  solutions of the field equations \eqref{LFeqt}$-$\eqref{GB_eq} with $\alpha'=0=\phi'$) have also been considered. 
For instance, in Ref. \cite{Ayon2000}, the Bardeen black hole solution  \cite{Bardeen1968} ~--the first regular black hole model in general relativity--~ was reinterpreted as a purely magnetic solution to Einstein
equations coupled to a nonlinear electrodynamics. The metric that describes the Bardeen black hole, in the framework of GR-NLED, is as follow 
\begin{equation}
ds^{2} =  - \left( 1 - \frac{2mr^{2}}{\sqrt{\left(r^{2} + q^{2}\right)^{3}}} \right) dt^{2} + \left( 1 - \frac{2mr^{2}}{\sqrt{\left(r^{2} + q^{2}\right)^{3}}} \right)^{-1}dr^{2}  + r^{2}d\Omega^{2}     
\end{equation}
where $q$ is the magnetic charge of a self-gravitating magnetic
field described by a nonlinear electrodynamics model, which depends only on the electromagnetic invariant $\mathcal{F}$, given by
\begin{equation}
 \mathcal{L}(\mathcal{F}) = \frac{3}{2sq^{2}}\!\left( \frac{ \sqrt{2q^{2}\mathcal{F}} }{1 + \sqrt{2q^{2}\mathcal{F}} }\right)^{\!\!^{\frac{5}{2}}}    
\end{equation}
where $s$ stands for $s=|q|/(2m)$, and for the NLED Lagrangian to be real-valued, $\mathcal{F}$ must be a positive quantity.
 
\section{\bf Rotation curves of spiral galaxies in E-STGB-NLED theory } 
According to \cite{Nucamendi2001,Harko10,Capozziello06}, assuming that stars and hydrogen gas clouds situated within the FRC zone of a spiral galaxy behave as test particles that follow the timelike geodesics of a SSS spacetime, 
it can be concluded that for an arbitrary SSS spacetime metric $ds^{2} = -e^{ A(r) }dt^{2} + e^{ B(r) }dr^{2}  + r^{2}d\Omega^{2}$ representing the gravitational field outside the galactic baryonic mass distribution (covering the FRC zone and the exterior of the spiral galaxy), the tangential velocity $v(r)$ of massive test particles (stars and/or neutral hydrogen gas clouds) in circular orbits (at radius $r$) around the galactic center
is independent of the function $B(r)$ and is given by $v(r)=\sqrt{rA'/2}$ (for details, see Refs. \cite{Weinberg,Landau}).  
Within the FRC zone, the velocity $v(r)$ remains approximately constant; nevertheless, observations \cite{Sanders1996}  
suggest that this constant value of $v(r)$ differs for each spiral galaxy\footnote{Here, the velocity is dimensionless since we use units where $c = 1$.}, usually ranging from $10^{-4}$ to $10^{-3}$.
Given these circumstances, the metric potential $A(r)$ in the FRC zone is obtained by integrating 
and takes the form 
\begin{equation}
A(r) \approx 2v^{2}\ln\!\left(\frac{r}{r_{g}}\right)    
\end{equation}
where $r_{g}$ is an integration constant with units of [length]. \\
On the other hand, it can be seen that 
given a NLED model determined by $\mathcal{L}(\mathcal{F})$
the following metric
\begin{equation}\label{ansatz}  
ds^{2} =  - e^{A(r)}dt^{2} + dr^{2}  + r^{2}d\Omega^{2}, \quad\textup{with}\quad  A(r) = \ln\!\left(\frac{\mathcal{F}}{\mathcal{L}(\mathcal{F})}\right), \quad\textup{being}\quad \mathcal{F} = \frac{q^{2}}{2r^{4}}  
\end{equation}
is an exact purely magnetic solution of the E-STGB-NLED field equations \eqref{LFeqt}$-$\eqref{GB_eq}, being $q$ the magnetic charge, with E-STGB gravity determined by the scalar field $\phi(r)=\int_{r_{\ast}}^{r} \phi'(\chi) d\chi$ with $\phi'(r)$ such that
\begin{equation}\label{sol_ph}
\phi'^{2}(r) = -\frac{4q^{2}}{r^{4}}e^{-A(r)}  
\end{equation}
and Gauss Bonnet coupling function $\alpha(r) =\alpha(\phi(r))$  
so that $\dot{\alpha}(r)=\dot{\alpha}(\phi(r))$ is given by 
\begin{equation}
\dot{\alpha}(r) = \frac{r}{4\phi'(r)} - \frac{r^{2}\phi'(r)}{8A'(r)} \quad\quad\textup{with}\quad\quad \phi'(r)\neq0
\end{equation}
hence $\alpha' = \phi'\dot{\alpha}$ becomes
\begin{equation}\label{sol_a}
 \alpha'(r) = \frac{r}{4} - \frac{r^{2}}{8A'(r)}\phi'^{2}(r).  
\end{equation}
Therefore, for this solution, the Gauss Bonnet coupling function in terms of the radial coordinate is obtained by integrating Eq. \eqref{sol_a}, {\it i.e.,} $\alpha(r)=\int_{r_{\ast}}^{r} \alpha'(\chi) d\chi$. \\
To prove this, we first note that for the metric \eqref{ansatz} the field equation \eqref{GB_eq} simplifies to the following equation:
\begin{equation}\label{GB_eq_pr}
r(\phi'^{2})' + (4 + rA')\phi'^{2} = 0
\end{equation}   
which is satisfied by \eqref{sol_ph}.
Therefore, once it is established that the field equations \eqref{EQ_gaa} are also satisfied, the proof will be complete.
Isolating $\mathcal{L}(\mathcal{F})$ from the expression for  
$A(r)$ presented in Eq. \eqref{ansatz} yields  
\begin{equation}\label{model_LF}
\mathcal{L}(\mathcal{F}) = \mathcal{F}e^{-A(r)} \quad\textup{  with  }\quad   
\quad  r = \left(\frac{q^{2}}{2\mathcal{F}} \right)^{\frac{1}{4}}    
\end{equation}
which in terms of the radial coordinate takes the form
\begin{equation}\label{sol_LF}  
\mathcal{L}(r)  =  \frac{q^{2}}{2r^{4}}e^{-A(r)}.   
\end{equation} 
Hence, for this solution,  $\mathcal{L}'(r)$  can be written as $\mathcal{L}'(r) = -\left[\frac{4}{r} + A'(r)\right]\!\mathcal{L}(r)$.  Thus,  $\mathcal{L}_{_{\mathcal{F}}}=\mathcal{L}'/\mathcal{F}'$ as function of $r$, becomes 
\begin{equation}\label{L_pri}
\mathcal{L}_{_{\mathcal{F}}}(r) = \frac{4 + r A'(r)}{4}e^{-A(r)}.      
\end{equation}
On the other hand, for this geometry 
the non null components of the Einstein tensor are
\begin{equation}\label{G_pr}
G_{t}{}^{t}= 0, \quad\quad G_{r}{}^{r}=\frac{A'}{r} \quad\quad \textup{ and } \quad\quad G_{\theta}{}^{\theta}=G_{\varphi}{}^{\varphi}=\frac{2A' + rA'^{2} + 2rA'' }{4r}.    
\end{equation}
The corresponding nontrivial components of $(T_{\mu}{}^{\nu})\!_{_{_{S\!G\!B}}}$  
for this solution are 
\begin{equation}\label{SGB_pr12}
8\pi(T_{t}{}^{t})\!_{_{_{S\!G\!B}}}  = \frac{q^{2}}{r^{4}}e^{-A(r)},
\quad\quad\quad 8\pi(T_{r}{}^{r})\!_{_{_{S\!G\!B}}} = \frac{A'}{r} + \frac{q^{2}}{r^{4}}e^{-A(r)} 
\end{equation}
and 
\begin{equation}\label{SGB_pr34} 
8\pi(T_{\theta}{}^{\theta})\!_{_{_{S\!G\!B}}} = 8\pi(T_{\varphi}{}^{\varphi})\!_{_{_{S\!G\!B}}} =  \frac{2A' + rA'^{2} + 2rA'' }{4r} - \frac{q^{2}}{2r^{4}}\!\left(2+rA'\right)\!e^{-A(r)}.  
\end{equation}
Whereas the nontrivial components of $(T_{\mu}{}^{\nu})\!_{_{_{N\!L\!E\!D}}}$, assuming that the electromagnetic field tensor is purely magnetic, are
\begin{equation}\label{NLED_pr12}
8\pi (T_{t}{}^{t})\!_{_{_{N\!L\!E\!D}}} = 8\pi (T_{r}{}^{r})\!_{_{_{N\!L\!E\!D}}} = -\frac{q^{2}}{r^{4}}e^{-A(r)},
\end{equation}
and 
\begin{equation}\label{NLED_pr34}
8\pi (T_{\theta}{}^{\theta})\!_{_{_{N\!L\!E\!D}}} = 8\pi (T_{\varphi}{}^{\varphi})\!_{_{_{N\!L\!E\!D}}} = \frac{q^{2}}{2r^{4}}\left(2+rA'\right)e^{-A(r)}  
\end{equation}
Therefore, Using \eqref{G_pr}$-$\eqref{NLED_pr34}, we can determine that the field equations $G_{\mu}{}^{\nu} = 8\pi(T_{\mu}{}^{\nu})\!_{_{_{S\!G\!B}}} + 8\pi (T_{\mu}{}^{\nu})\!_{_{_{N\!L\!E\!D}}}$ are satisfied. \\
Up to this point, we have made no assumptions about the form of the NLED Lagrangian; hence, the above holds for any function $\mathcal{L}(\mathcal{F})$ of class $\mathcal{C}^{1}$.
Nevertheless, whenever the NLED model $\mathcal{L}(\mathcal{F})$ in the weak-field limit reduces to linear electromagnetic theory (also know as Maxwell weak field limit; $\mathcal{L}(\mathcal{F}) \sim \mathcal{F}$ and $\mathcal{L}_{_{\mathcal{F}}}(\mathcal{F})\sim 1$ as $\mathcal{F}\sim0$), Eq. \eqref{ansatz} produces  an asymptotically flat metric (that is, the resulting metric is such that; $ds^{2} \sim - dt^{2} + dr^{2}  + r^{2}d\Omega^{2} ~\textup{ as }~ r\sim\infty$).
On the other hand, an essential limitation on the form of the NLED Lagrangian $\mathcal{L}(\mathcal{F})$ 
arises when we take into account that a matter distribution represented by an energy-momentum tensor
$T_{\mu\nu}$ is considered physically reasonable classical matter when it satisfies the energy conditions, specifically the null energy condition (NEC) and the weak energy condition (WEC). 
More precisely, for a matter distribution, the fulfillment of NEC and WEC ensures that the quantity $T_{\mu\nu}\xi^{\mu}\xi^{\nu}$, which physically represents the energy density of matter distribution as measured by an observer whose 4$-$velocity is $\boldsymbol{\xi}=\xi^{\nu}\partial_{\nu}$ (where $\boldsymbol{\xi}$ is any null vector for the NEC, whereas $\boldsymbol{\xi}$ is any timelike vector for the WEC), is non-negative (for review, see \cite{Wald}).
Therefore, for the purely magnetic SSS configurations, NLED Lagrangian $\mathcal{L}(\mathcal{F})$ 
satisfies the NEC and WEC as long as $\mathcal{L}(\mathcal{F})$ and its derivative  
satisfy specific positivity constraints given by: $\mathcal{L}(\mathcal{F})\geq0$ and $\mathcal{L}_{_{\mathcal{F}}}(\mathcal{F})\geq0$ everywhere in spacetime (for a recent review, see \cite{Canate23,Canate21}). 
Nevertheless, taking into account equations \eqref{sol_ph} and \eqref{sol_LF}, we arrive at the following expression
\begin{equation}\label{sol_phr}
\phi'^{2}(r) = - 8\mathcal{L}(r)      
\end{equation}
which indicates that $\phi(r)$ becomes a phantom scalar field whenever the NLED Lagrangian $\mathcal{L}(\mathcal{F})$ fulfills the null and weak energy conditions. In other words, for the purely magnetic E-STGB-NLED solution \eqref{ansatz}$-$\eqref{sol_a}, as long as $\mathcal{L}(\mathcal{F})>0$ yields $\phi(r)=\int^{r}_{r_{g}}\phi'(r)dr\in\mathbb{I}$. However, it can be deduced that in this instance,
since Eq. \eqref{sol_ph} states that $\phi'^{2}(r)$ is scaled by $q^{2}$, small values of $q^{2}$ can arbitrarily minimize the amount of exotic matter (phantom scalar field) included in the SGB tensor associated with the solution \eqref{ansatz}$-$\eqref{sol_a}.

\subsection{Nonlinear electrodynamics model extremely close 
to Maxwell electrodynamics} 
The main result of this work arises from considering the following NLED Lagrangian defined by: 
\begin{equation}\label{NLEDmodel}
\mathcal{L}(\mathcal{F}) = \left\{ \frac{\left[\left(\mathcal{F}_{_{0}} \right)^{-\frac{1}{4}}
- \left(\mathcal{F}_{_{s}} \right)^{-\frac{1}{4}} \right]\left(\mathcal{F} \right)^{-\frac{3}{4}} }{~\left[\left(\mathcal{F}_{_{0}} \right)^{-\frac{1}{4}} - \left(\mathcal{F}_{_{s}} \right)^{-\frac{1}{4}}\right] \left(\mathcal{F}\right)^{-\frac{3}{4}} + \left(\mathcal{F}_{_{s}} \right)^{-\frac{1}{4}} \left(\mathcal{F}_{_{0}} \right)^{-\frac{3}{4}}~}\beta(\mathcal{F}) + \left(\frac{\mathcal{F}_{_{g}}}{\mathcal{F}}\right)^{\!\!\frac{v^{2}}{2}}\!\sigma(\mathcal{F}) \right\}^{\!\!\!^{-1}}\mathcal{F}
\end{equation}
being $\sigma(\mathcal{F})$ a smooth function given by:
\begin{equation}\label{sigma1}
\sigma(\mathcal{F}) = \frac{1}{2}  - \frac{1}{2} \tanh\!\left(\frac{\mathcal{F}_{_{g}}}{\mathcal{F}} - \frac{\mathcal{F}_{_{g}}}{\mathcal{F}_{_{c}}}\right) 
\end{equation} 
such that $0\leqslant\sigma(\mathcal{F})\leqslant 1$ for all $\mathcal{F}\in\mathbb{R}$, 
whereas $\beta(\mathcal{F}) = 1-\sigma(\mathcal{F})$.
The  quantities $\{ \mathcal{F}_{_{0}}, \mathcal{F}_{_{s}}, \mathcal{F}_{_{g}}, \mathcal{F}_{_{c}}\}$ 
are parameters that determine the scale of the NLED model and correspond to specific values of the electromagnetic invariant $\mathcal{F}$, while $v$ is a dimensionless constant. \\ 
There are two options to keep the NLED Lagrangian real: 
(i) all the parameters $\{ \mathcal{F}_{_{0}}, \mathcal{F}_{_{s}}, \mathcal{F}_{_{g}}, \mathcal{F}_{_{c}}\}$ and the invariant $\mathcal{F}$ are positive quantities; or (ii) all the parameters  $\{ \mathcal{F}_{_{0}}, \mathcal{F}_{_{s}}, \mathcal{F}_{_{g}}, \mathcal{F}_{_{c}}\}$ and the invariant $\mathcal{F}$ are negative quantities.
However, in the remainder of this paper, only option (i) will be considered.\\
For this option, in order for the NLED Lagrangian \eqref{NLEDmodel} to be positive-valued and asymptotically Maxwell in both weak and strong field limits\footnote{ For option (ii), $\mathcal{F}\in(-\infty ,0]$, asymptotically Maxwell in both weak and strong field limits means: $\mathcal{L}(\mathcal{F})\approx\mathcal{F}$ as $\mathcal{F}\sim0$ and as $\mathcal{F}\sim-\infty$. } (that is, $\mathcal{L}(\mathcal{F})\approx\mathcal{F}$ as $\mathcal{F}\sim0$ and as $\mathcal{F}\sim\infty$), the following conditions must be satisfied: 
\begin{equation} \label{CND}
0\leqslant v \ll 1 
\quad\quad\textup{ and }\quad\quad
\mathcal{F}_{_{s}} \gg \mathcal{F}_{_{g}} \gg \mathcal{F}_{_{c}} \geqslant \mathcal{F}_{_{0}} > 0.
\end{equation}
In fact, as long as the conditions \eqref{CND} are satisfied, our NLED model   
will closely approximate Maxwell electrodynamics, that is:
\begin{equation}
\mathcal{L}(\mathcal{F}) \approx \mathcal{F} \quad\quad\textup{for}\quad\quad \mathcal{F}\in [0,\mathcal{F}_{_{c}}-\epsilon^{2})\cup(\mathcal{F}_{_{c}}+\epsilon^{2},\infty)    
\end{equation}
where $\epsilon^{2}$ represents a small positive number that can be arbitrarily close to zero depending on the choice of $\mathcal{F}_{_{g}}$ and $\mathcal{F}_{_{c}}$, so that $\epsilon^{2}\sim0$ as long as $\frac{\mathcal{F}_{_{c}}}{\mathcal{F}_{_{g}}}\sim 0$. \\
{\it Proof:} Let us start with the assumption that 
the conditions \eqref{CND} are fulfilled.  In this case, 
the function $\sigma(\mathcal{F})$ is such that
\begin{equation}\label{sigma1Bh}
\sigma(\mathcal{F}) \approx 0 \quad\quad\textup{for}\quad\quad \mathcal{F} \in [0,\mathcal{F}_{_{c}}-\epsilon^{2}) \quad\quad\textup{and}\quad\quad  \sigma(\mathcal{F}) \approx 1 \quad\quad\textup{for}\quad\quad \mathcal{F} \in (\mathcal{F}_{_{c}}+\epsilon^{2},\infty)   
\end{equation}
where $\epsilon^{2}$ represents a small positive number that can be arbitrarily close to zero depending on the choice of $\mathcal{F}_{_{g}}$ and $\mathcal{F}_{_{c}}$, 
so that $\epsilon^{2}\sim0$ as long as $\frac{\mathcal{F}_{_{c}}}{\mathcal{F}_{_{g}}}\sim 0$.
Consequently, for $\mathcal{F}\in[0,\mathcal{F}_{_{c}}-\epsilon^{2})$, the NLED model \eqref{NLEDmodel} can be approximated as
\begin{equation}\label{out_NLED}
\mathcal{L}(\mathcal{F}) \approx  
\frac{~\left[\left(\mathcal{F}_{_{0}} \right)^{-\frac{1}{4}} - \left(\mathcal{F}_{_{s}} \right)^{-\frac{1}{4}}\right] \left(\mathcal{F}\right)^{-\frac{3}{4}} + \left(\mathcal{F}_{s} \right)^{-\frac{1}{4}} \left(\mathcal{F}_{_{0}} \right)^{-\frac{3}{4}}}{\left[\left(\mathcal{F}_{_{0}} \right)^{-\frac{1}{4}}
- \left(\mathcal{F}_{_{s}} \right)^{-\frac{1}{4}} \right]\left(\mathcal{F} \right)^{-\frac{3}{4}} }~
\mathcal{F}
\end{equation}
which,  given that $0 < \frac{\mathcal{F}_{_{0}}}{\mathcal{F}_{_{s}}} \ll \frac{\mathcal{F}_{_{0}}}{\mathcal{F}_{_{c}}} \leqslant 1$ according to    \eqref{CND}, becomes approximately equal to
\begin{equation}
\frac{~\left[\left(\mathcal{F}_{_{0}} \right)^{-\frac{1}{4}} - \left(\mathcal{F}_{_{s}} \right)^{-\frac{1}{4}}\right] \left(\mathcal{F}\right)^{-\frac{3}{4}} + \left(\mathcal{F}_{s} \right)^{-\frac{1}{4}} \left(\mathcal{F}_{_{0}} \right)^{-\frac{3}{4}}}{\left[\left(\mathcal{F}_{_{0}} \right)^{-\frac{1}{4}}
- \left(\mathcal{F}_{_{s}} \right)^{-\frac{1}{4}} \right]\left(\mathcal{F} \right)^{-\frac{3}{4}} } = \frac{~\left[ 1 - \left(\frac{\mathcal{F}_{_{0}}}{\mathcal{F}_{_{s}}} \right)^{\frac{1}{4}}\right] \left(\frac{\mathcal{F}_{_{0}}}{\mathcal{F}}\right)^{\frac{3}{4}} + \left( \frac{\mathcal{F}_{_{0}}}{ \mathcal{F}_{_{s}} } \right)^{\frac{1}{4}}  }{\left[1
- \left(\frac{\mathcal{F}_{_{0}}}{\mathcal{F}_{_{s}}} \right)^{\frac{1}{4}} \right]\left(\frac{\mathcal{F}_{_{0}}}{\mathcal{F}}\right)^{\frac{3}{4}} }   
\approx 1 \quad\textup{~for~}\quad \mathcal{F}\in[0,\mathcal{F}_{_{c}}-\epsilon^{2}) 
\end{equation}
and then 
\begin{equation}\label{ext_L}
\mathcal{L}(\mathcal{F}) \approx \mathcal{F} \quad\textup{~for~}\quad \mathcal{F}\in[0,\mathcal{F}_{_{c}}-\epsilon^{2}).
\end{equation}
While, for $\mathcal{F}\in(\mathcal{F}_{_{c}}+\epsilon^{2},\infty)$ the NLED model \eqref{NLEDmodel} can be approximated as 
\begin{equation}\label{ins_NLED}
\mathcal{L}(\mathcal{F}) \approx   
\left(\frac{\mathcal{F}}{\mathcal{F}_{_{g}}}\right)^{\!\!1+\frac{v^{2}}{2}}~\mathcal{F}_{_{g}}
\end{equation}
which, since $0\leq v\ll1$   
produces $1+\frac{v^{2}}{2} \approx 1$, and then  
\begin{equation}\label{int_L}
\mathcal{L}(\mathcal{F}) \approx \mathcal{F} \quad\textup{~for~}\quad \mathcal{F}\in(\mathcal{F}_{_{c}}+\epsilon^{2},\infty).
\end{equation}
Therefore, Eqs. \eqref{ext_L} and \eqref{int_L} result in $\mathcal{L}(\mathcal{F}) \approx \mathcal{F}$ for $\mathcal{F}\in [0,\mathcal{F}_{_{c}}-\epsilon^{2})\cup(\mathcal{F}_{_{c}}+\epsilon^{2},\infty)$. 

\subsection{Purely magnetic spiral galaxy spacetime in E-STGB-NLED gravity} 
\begin{figure}[H] 
    \centering
    \includegraphics[width=0.3\textwidth]{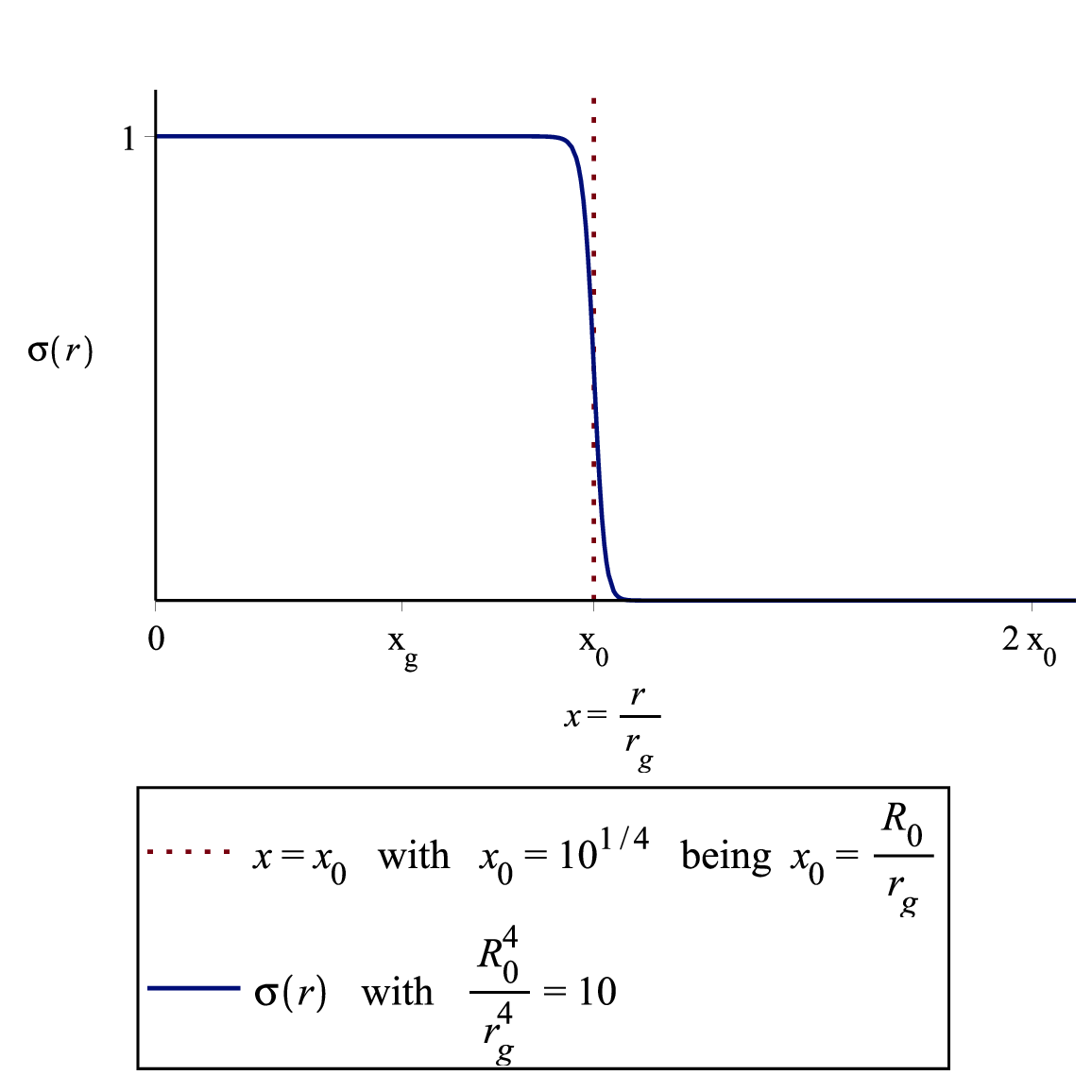}
    \hfill
    \includegraphics[width=0.3\textwidth]{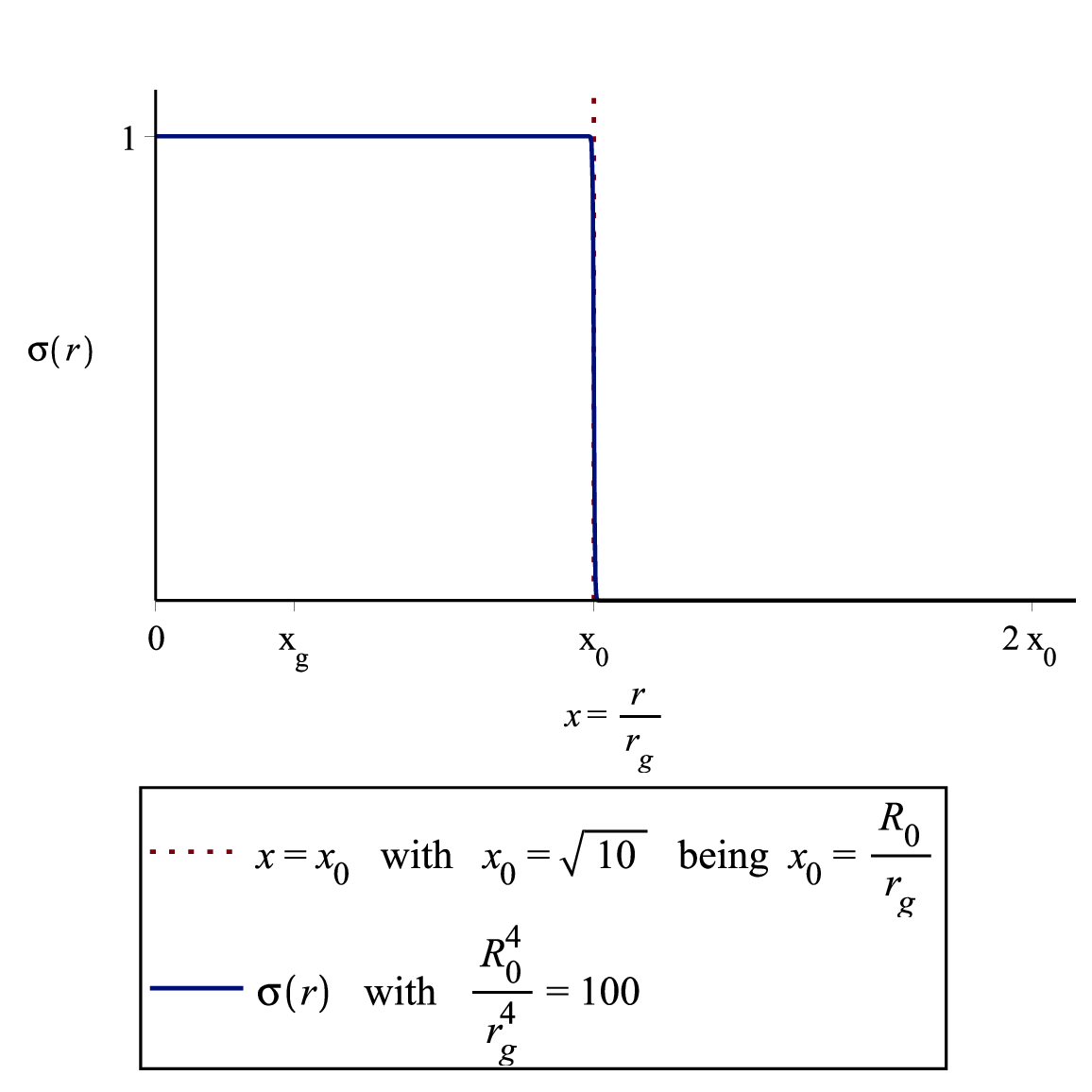}
    \hfill
    \includegraphics[width=0.3\textwidth]{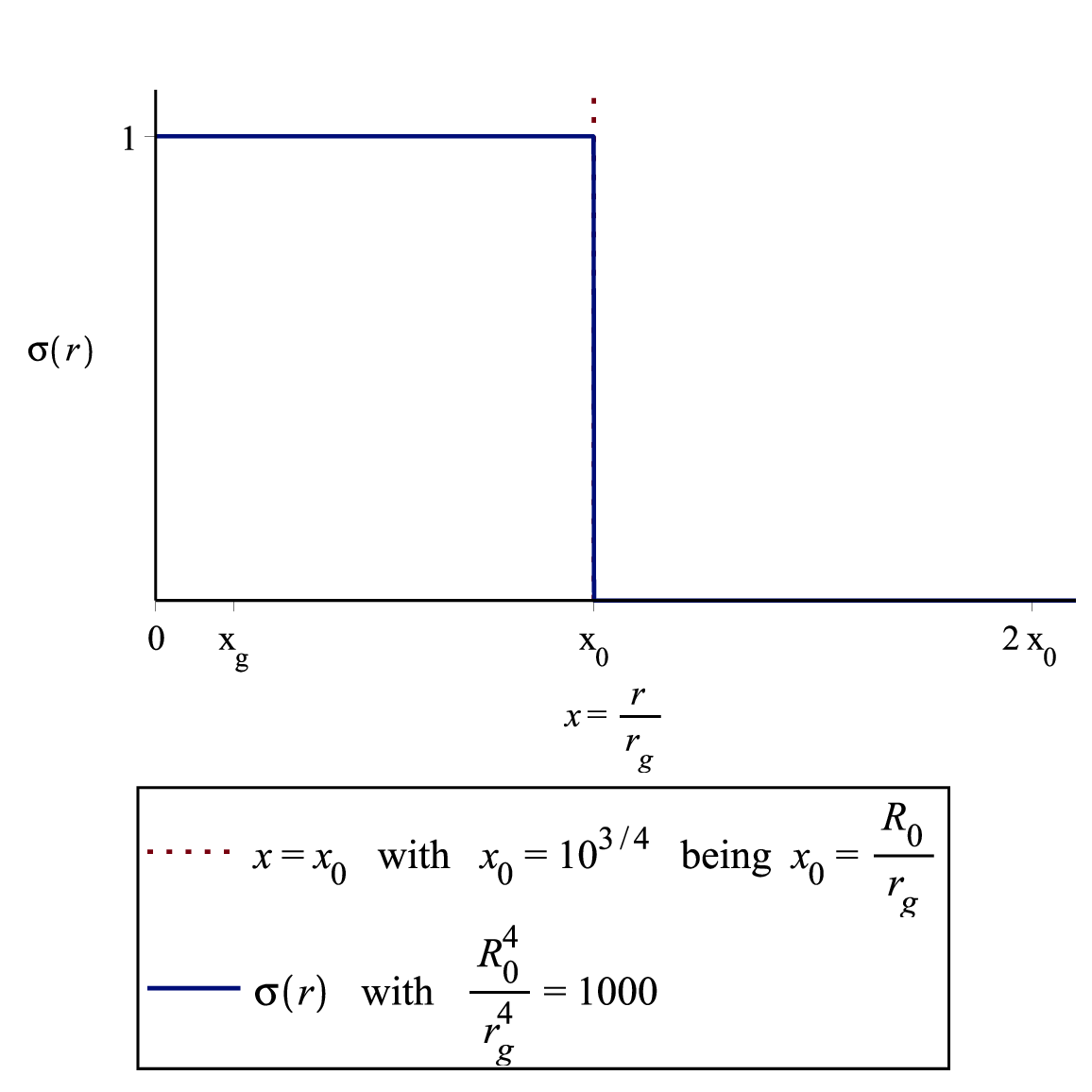}
    \caption{\label{FIG1} Behavior of $\sigma(r)$ with respect to the scaled variable $x= \frac{r}{r_{_{g}}}$
    for different values of the ratio $\frac{R_{0}}{r_{g}}$. Solid line (\textcolor{blue}{$\rule[0.8mm]{6mm}{0.3mm}$}) represents the function $\sigma(r)$, while dot line (\textcolor{Brown}{$\cdot\!~\!\cdot\!~\!\cdot\!~\!\cdot\!~\!\cdot$}) represents the region $x = x_{_{0}} = \frac{R_{_{0}}}{r_{_{g}}}$.
    Left: Plots with $\frac{R_{_{0}}}{r_{_{g}}}=10^{^{\frac{1}{4}}}$. Middle: Plots with $\frac{R_{_{0}}}{r_{_{g}}}=10^{^{\frac{1}{2}}}$. Right: Plots with $\frac{R_{_{0}}}{r_{_{g}}}=10^{^{\frac{3}{4}}}$.
    }
\end{figure}

Since for the purely magnetic SSS case $\mathcal{F}$ takes the form $\mathcal{F}=\frac{q^{2}}{2r^{4}}\in\mathbb{R}^{+}+\{0\}$, we renamed 
$\mathcal{F}_{_{0}}=q^{2}/(2r^{4}_{_{0}})$, $\mathcal{F}_{_{s}}=q^{2}/(2\mu^{4})$, $\mathcal{F}_{_{g}}=q^{2}/(2r^{4}_{_{g}})$ and $\mathcal{F}_{_{c}}=q^{2}/(2R^{4}_{_{0}})$.   
Then according to Eq. \eqref{ansatz}, in the context of E-STGB-NLED gravity, we can conclude that 
the NLED model \eqref{NLEDmodel} can be interpreted as the field source of the following purely magnetic AF-SSS spacetime metric: 
\begin{equation}\label{RCG}
ds^{2} =  - \left[ \frac{(r_{0}-\mu)r^{3} }{~(r_{0}-\mu)r^{3} + \mu r^{3}_{0}~}\beta(r) + \left(\frac{r}{r_{g}}\right)^{\!\!2v^{2}}\!\sigma(r) \right]dt^{2} + dr^{2}  + r^{2}d\Omega^{2}, \quad\quad\textup{ for }\quad\quad  r \in (0,\infty)
\end{equation}
being $\beta(r)=1-\sigma(r)$ where $\sigma(r)$, according to Eq. \eqref{sigma1}, is given by: 
\begin{equation}\label{sigma1r}
\sigma(r)  = \frac{1}{2} - \frac{1}{2}\tanh\left[\left(\frac{r}{r_{_{g}}}\right)^{4}-\left(\frac{R_{_{0}}}{r_{_{g}}}\right)^{4}\right]       
\end{equation}
with  scalar field $\phi(r)=\int_{r_{\ast}}^{r} \phi'(\chi) d\chi$ and GB coupling function $\alpha(\phi(r))=\int_{r_{\ast}}^{r} \alpha'(\chi) d\chi$
determined, respectively, by \eqref{sol_ph} and \eqref{sol_a}.
The parameters $r_{_{0}}=\left[q^{2}/(2\mathcal{F}_{_{0}})\right]^{1/4}$, $R_{_{0}}=\left[q^{2}/(2\mathcal{F}_{_{c}})\right]^{1/4}$, $r_{_{g}}=\left[q^{2}/(2\mathcal{F}_{_{g}})\right]^{1/4}$ and $\mu=\left[q^{2}/(2\mathcal{F}_{_{s}})\right]^{1/4}$  are positive with units of [length] and, since \eqref{CND}, must satisfy:
\begin{equation}\label{CNDr}
r^{4}_{_{0}} \geqslant  R^{4}_{_{0}} \gg r^{4}_{_{g}} \gg \mu^{4}.
\end{equation}
Hence, the behavior of $\sigma(r)$ Eq. \eqref{sigma1r} is such that;
$\sigma(r) \approx 1$ at $r\in [0,R_{_{0}}-\delta^{2})$, and $\sigma(r) \approx 0$ at $r\in (R_{_{0}}+\delta^{2},\infty)$, where $\delta^{2}$ represents an arbitrarily small positive number so that $\delta^{2} \sim 0$ as $\frac{r_{_{g}}}{R_{_{0}}} \sim 0$. See Fig. \ref{FIG1}.  
As a result, in the context of spiral galaxies, the metric \eqref{RCG} with parameters that satisfy condition \eqref{CNDr} and defining $R_{_{0}}$ as the radius where the flat rotation curves end, at $r\in (0,R_{_{0}}-\delta^{2})$, can be expressed approximately as:
\begin{equation}\label{ins_met}
ds^{2} \approx  - \left(\frac{r}{r_{_{g}}}\right)^{2v^{2}}dt^{2} + dr^{2}  + r^{2}d\Omega^{2}, \quad\quad\textup{ for }\quad\quad  r \in (0,R_{_{0}}-\delta^{2})    
\end{equation}
describing the  gravitational field within the FRC zone of a spiral galaxy, setting $v= \textup{constant} \in[10^{-4}\!~,~\!10^{-3}]$  
based on the observations; whereas at $r\in(R_{_{0}}+\delta^{2},\infty)$, can be expressed approximately as: 
\begin{equation}\label{out_met}
ds^{2} \approx  - \frac{(r_{_{0}}-\mu)r^{3}}{(r_{_{0}}-\mu)r^{3} + \mu r^{3}_{_{0}}}dt^{2} + dr^{2}  + r^{2}d\Omega^{2}, \quad\quad\textup{ for }\quad\quad  r \in (R_{_{0}}+\delta^{2},\infty)   
\end{equation}
which describes an AF spacetime that represents the exterior of the spiral galaxy, where $r_{_{0}}$ indicates the radial distance of closest approach to the center of the galaxy for a photon traveling a very long distance toward the spiral galaxy. Whereas, as discussed below, $\mu$ is connected to the mass of the spiral galaxy. \\
An alternative model of $\sigma(\mathcal{F})$ that produces a similar meaning when applied to \eqref{RCG} is provided by:
\begin{equation}\label{sigma2F}
\sigma(\mathcal{F}) = \frac{1}{2}\tanh\left(\frac{\mathcal{F}_{_{g}}}{\mathcal{F}} + \frac{\mathcal{F}_{_{g}}}{\mathcal{F}_{_{c}}}\right)-\frac{1}{2}\tanh\left(\frac{\mathcal{F}_{_{g}}}{\mathcal{F}} - \frac{\mathcal{F}_{_{g}}}{\mathcal{F}_{_{c}}}\right)  
\end{equation}
which (for the purely magnetic SSS case) in terms of $r$ has the form
\begin{equation}\label{sigma2r}
\sigma(r) = \frac{1}{2}\tanh\left(\frac{ r^{4}+R^{4}_{_{0}} }{r^{4}_{_{g}}}\right) - \frac{1}{2}\tanh\left( \frac{ r^{4}-R^{4}_{_{0}} }{r^{4}_{_{g}}} \right)
\end{equation}
The above function behaves similarly to $\sigma(r)$ given by Eq. \eqref{sigma1r} if the condition\footnote{Notice that the function $\sigma(r)$ given by Eq. \eqref{sigma1r}, or by  Eq. \eqref{sigma2r}, with $r_{_{g}},R_{_{0}}\in\mathbb{R}^{+}-\{0\}$, for very small values of $\frac{r_{_{g}}}{R_{_{0}}}$ ({\it i.e.,} as $\frac{r_{g}}{R_{0}}\approx0$), very quick approaches a top hat function; $\sigma(r)\approx1$ at $r\in [0, R_{_{0}})$, and $\sigma(r)\approx0$ at $r\in (R_{_{0}},\infty)$.} \eqref{CNDr} is satisfied.
{\it I.e.,}  $\sigma(r) \approx 1$ at $r\in [0,R_{_{0}}-\delta^{2})$, and $\sigma(r) \approx 0$ at $r\in (R_{_{0}}+\delta^{2},\infty)$, being $\delta^{2}$ a positive number such that $\delta^{2} \sim 0$ as $\frac{r_{_{g}}}{R_{_{0}}} \sim 0$. \\
On the other hand, in the framework of general relativity, the Birkhoff theorem implies that the metric \eqref{RCG} cannot be presented as a vacuum solution in GR.  
Additional matter fields are necessary to support this kind of spacetime as a solution of the Einstein field equations. 
Nevertheless, as we demonstrate below, the metric \eqref{RCG} cannot be presented as a purely magnetic solution of the GR-NLED-SF field equations. \\
In the framework of general relativity minimally coupled to a free scalar field (GR-SF)
for SSS spacetime, {\it i.e.,} the field equations \eqref{LFeqt}$-$\eqref{GB_eq} with  
$\boldsymbol{F} =0$, $\mathcal{F}=0=\mathcal{L}(\mathcal{F})$ and $\alpha'(r)=0$,
if we consider a spacetime metric of the form 
\begin{equation}\label{A_metric}
ds^{2} =  - e^{A(r)}dt^{2} + dr^{2}  + r^{2}d\Omega^{2}    
\end{equation} 
the Eq. \eqref{LFeqt} implies that $\phi'(r)=0$ for all $r$. Eq. \eqref{LFeqr} yields $A(r)=0$, whereas Eqs. \eqref{LFeqphi}$-$\eqref{GB_eq} becomes trivial $0=0$. As a result, the metric ansatz \eqref{A_metric} becomes the Minkowski metric. 
We can therefore conclude that the spacetime metric \eqref{RCG} is not permitted as a solution of GR-SF field equations. \\
On the other hand, in the framework of GR-NLED for purely magnetic SSS spacetime, {\it i.e.,} the equations \eqref{LFeqt}$-$\eqref{GB_eq} with $\phi'(r)=0=\alpha'(r)$ and electromagnetic field tensor given by \eqref{magnetica},
if we consider the metric ansatz Eq. \eqref{A_metric}, 
the Eq. \eqref{LFeqt} shows that the electromagnetic source vanishes, that is $\mathcal{L}(\mathcal{F})=0$ for all $r$. 
Eq. \eqref{LFeqr} yields $A(r)=0$, while Eqs. \eqref{LFeqphi}$-$\eqref{GB_eq} becomes trivial $0=0$. 
Consequently, the only possible solution is the Minkowski metric. 
Hence the spacetime metric \eqref{RCG} cannot be interpreted as a purely magnetic solution of GR-NLED field equations.  \\
On the other hand, in the context of GR-NLED-SF theory for purely magnetic SSS spacetime, {\it i.e.,} the field equations \eqref{LFeqt}$-$\eqref{GB_eq} with $\alpha'(r)=0$ and electromagnetic field tensor given by \eqref{magnetica}, we find that there is only one NLED model that can produce a purely magnetic GR-NLED-SF solution consistent with the metric ansatz \eqref{A_metric}, and this model is defined by the NLED Lagrangian: 
\begin{equation}\label{Lsol}
\mathcal{L}(\mathcal{F})=\frac{\eta\!~\mathcal{F}}{1 + \eta\!~\sqrt{2q^{2}\mathcal{F}} }
\end{equation}
where $\eta$ is a parameter dimensionless of the model. This special purely magnetic GR-NLED-SF solution is determined by the following metric:
\begin{equation}\label{no_GxM}
ds^{2} =  -  \left(\frac{r^{2} + \eta\!~ q^{2}}{\eta\!~r^{2}}\right)dt^{2} + dr^{2}  + r^{2}d\Omega^{2}    
\end{equation}
and a massless scalar field $\phi(r)=\int_{r_{\ast}}^{r}\phi'(\chi) d\chi$, where $\phi'(r)$ is determined by: 
\begin{equation}
\phi'^{2}(r) = -\frac{ 4\!~\eta\!~q^{2}}{(r^{2}+ \eta\!~q^{2})r^{2}}. 
\end{equation}
Nonetheless, since the metric \eqref{no_GxM} does not belong to the type \eqref{RCG},
the spacetime metric \eqref{RCG} cannot be interpreted as a purely magnetic solution of GR-NLED-SF field equations.
\\
Therefore, the metric given by Eq. \eqref{RCG} which describes the gravitational field of a spiral galaxy, becomes an interesting type of purely magnetic spacetime  that arises when considering E-STGB-NLED gravity.
For this type of spacetime geometry, in addition to the previously mentioned features, the Schwarzschild metric results on solar system scales (such as the deflection of light from the sun) must be recovered. To accomplish this, we will specify the parameters $\mu$ and $r_{0}$ as demonstrated below. 

\subsection{GRAVITATIONAL LENSING EFFECTS}
\begin{figure}[H]
    \centering
    \includegraphics[width=0.35\textwidth]{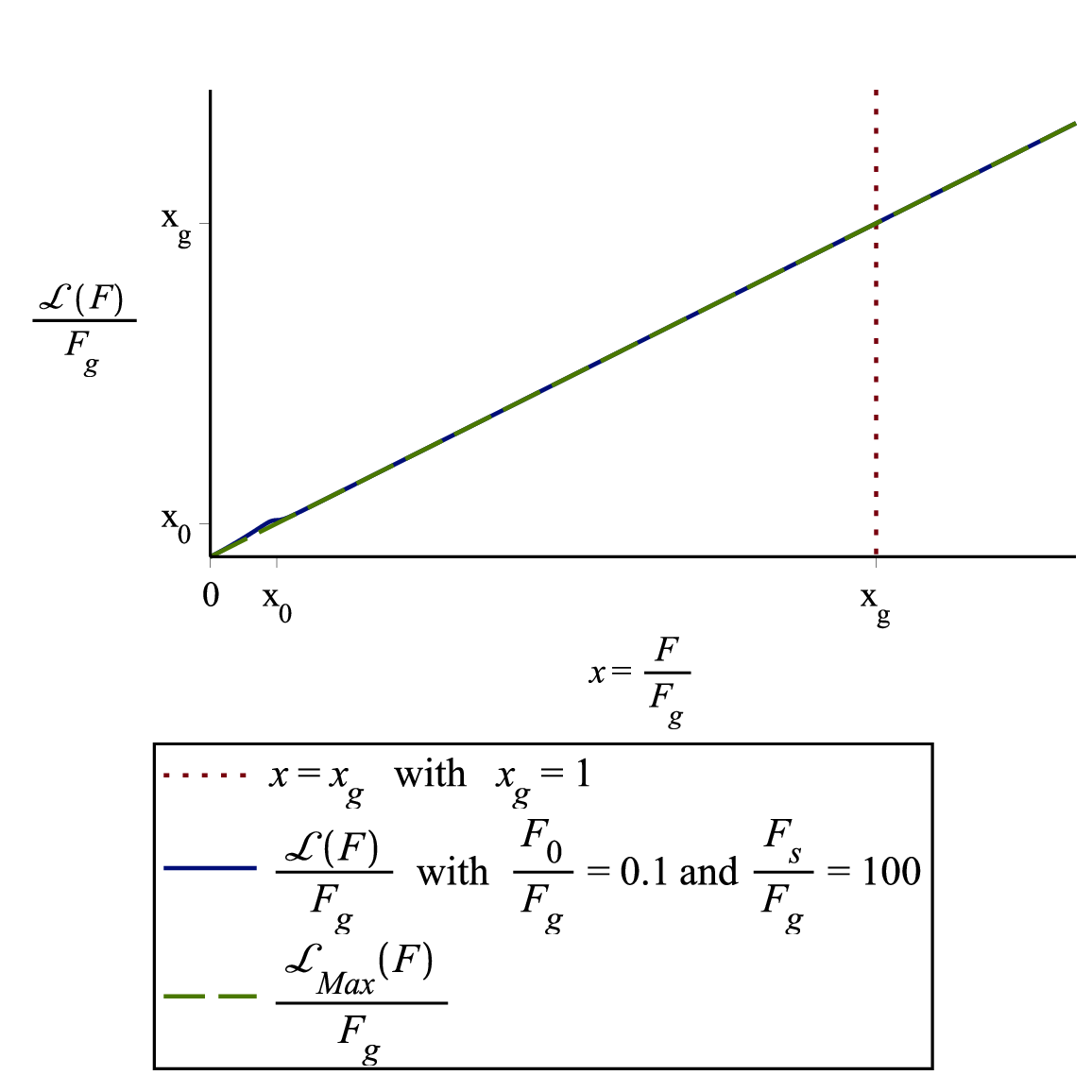}
    \hfill 
    \includegraphics[width=0.35\textwidth]{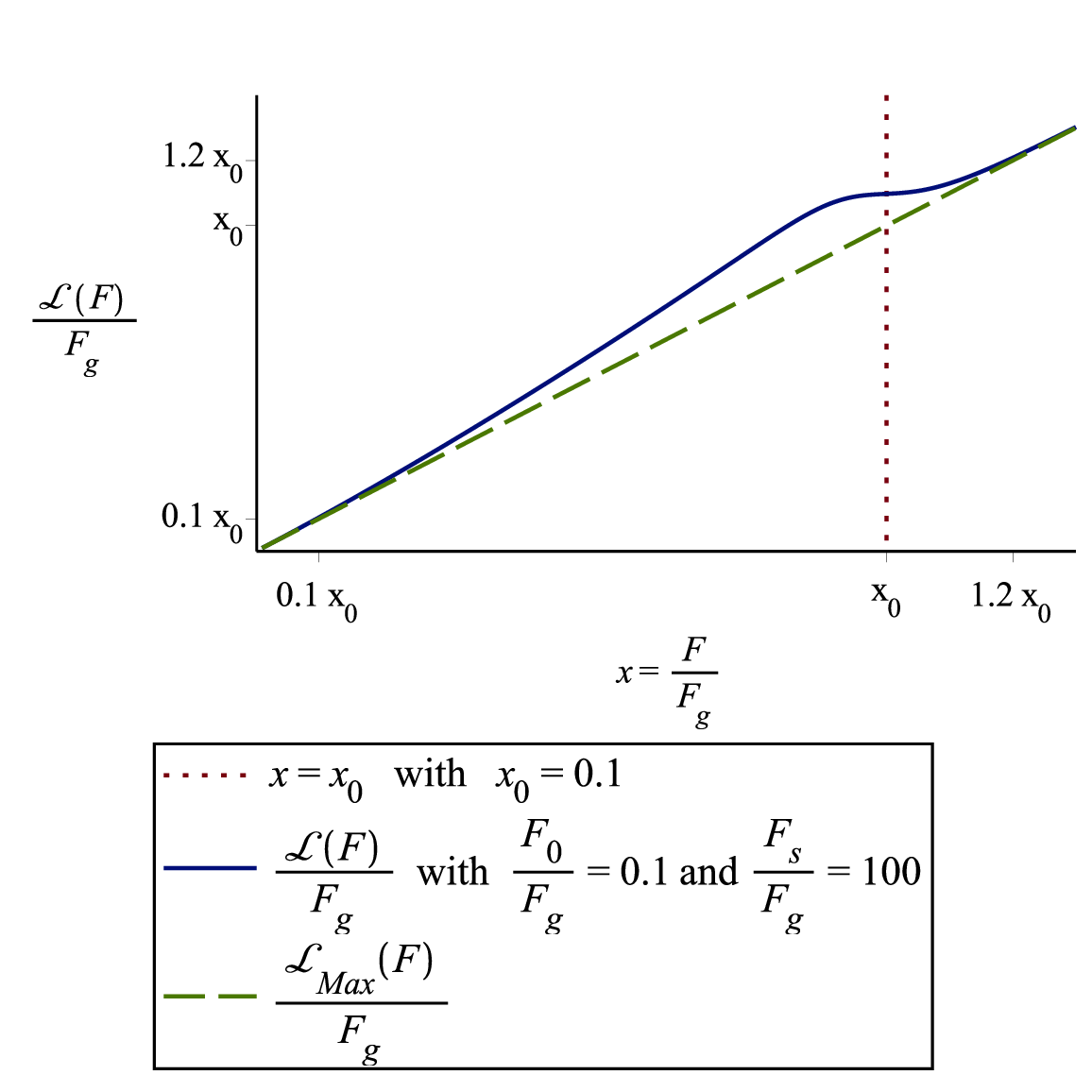}
    \caption{\label{FIG2} The dimensionless electromagnetic Lagrangian $\frac{\mathcal{L}(\mathcal{F})}{\mathcal{F}_{_{g}}}$ versus the dimensionless electromagnetic invariant $x=\frac{\mathcal{F}}{\mathcal{F}_{_{g}}}$ is plotted.
    Solid line (\textcolor{blue}{$\rule[0.8mm]{6mm}{0.3mm}$}) represents the 
    dimensionless Lagrangian for the NLED model \eqref{NLEDmodel} with $\mathcal{F}_{_{c}}=\mathcal{F}_{_{0}}$, $x_{_{0}}=\frac{\mathcal{F}_{_{0}}}{\mathcal{F}_{_{g}}} = 10^{-1}$, $x_{_{s}} = \frac{\mathcal{F}_{_{s}}}{\mathcal{F}_{_{g}}} = 10^{2}$, and the plot essentially stays the same regardless of the value of the constant $v$ fixed as $v\in[10^{-4}\!~,~\!10^{-3}]$ in accordance with the issue of flattening spiral galaxy rotation curves. For this NLED configuration, the parameters ($R_{_{0}}$, $r_{_{g}}$, $M$, $v$) which determine the solution \eqref{RCG_D},  are such that $R^{4}_{_{0}}= 10~\!r^{4}_{_{g}}$, $r^{4}_{_{g}}= 10^{2}~\!(2M)^{4}$ and $v = \textup{constant} \in[10^{-4}\!~,~\!10^{-3}]$,  fulfilling  \eqref{CNDrSG}.
    Dashed line (\textcolor{OliveGreen}{$\rule[0.8mm]{3.0mm}{0.3mm}~\rule[0.8mm]{3.0mm}{0.3mm}$}) represents the
    dimensionless Lagrangian for the Maxwell electrodynamics theory. Left panel shows plots in the interval $x=\frac{\mathcal{F}}{\mathcal{F}_{_{g}}}\in[0,1.3~\!x_{_{g}})$, dot line (\textcolor{Brown}{$\cdot\!~\!\cdot\!~\!\cdot\!~\!\cdot\!~\!\cdot$}) represents the region $x = x_{_{g}} =\frac{\mathcal{F}_{_{g}}}{\mathcal{F}_{_{g}}}= 1$. Right panel shows plots in the interval $x=\frac{\mathcal{F}}{\mathcal{F}_{_{g}}}\in(0.01~\!x_{_{0}},1.3~\!x_{_{0}})$, dot line (\textcolor{Brown}{$\cdot\!~\!\cdot\!~\!\cdot\!~\!\cdot\!~\!\cdot$}) represents the region $x = x_{_{0}} =\frac{\mathcal{F}_{_{0}}}{\mathcal{F}_{_{g}}} = 10^{-1}$.
    These plots indicate that the NLED Lagrangian \eqref{NLEDmodel}, for this setting of parameters, is monotonically increasing and positive in the domain $\mathcal{F}\in[0,\infty)$. Furthermore, in this instance, it can be noted that over the whole domain $\mathcal{F}\in[0,\infty)$, with the exception of a small region around $\mathcal{F}\approx\mathcal{F}_{_{0}}$, the NLED Lagrangian \eqref{NLEDmodel} is surprisingly similar to the Maxwell Lagrangian. }
\end{figure}

\begin{figure}[H]
    \centering
    \includegraphics[width=0.35\textwidth]{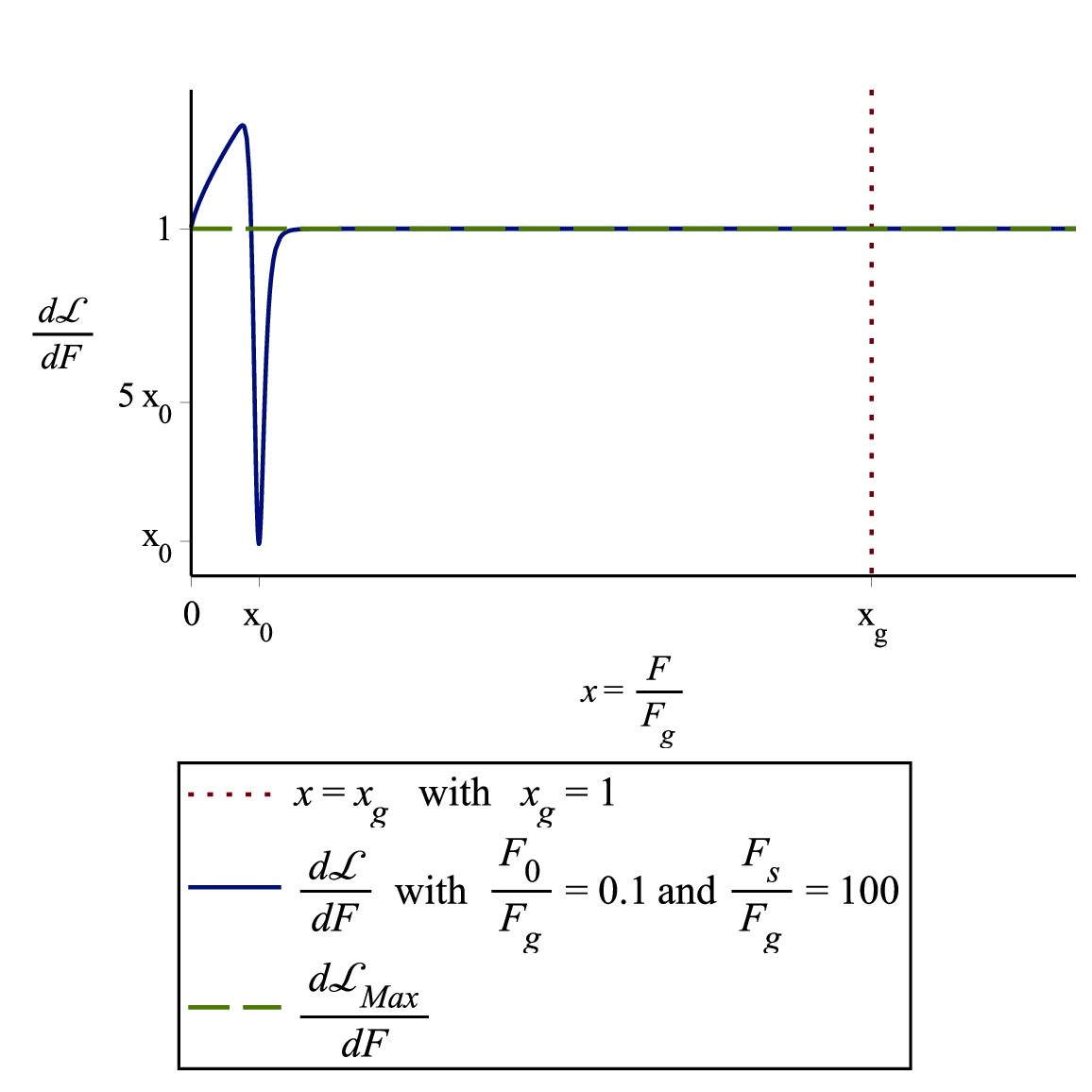}
    \hfill %
    \includegraphics[width=0.35\textwidth]{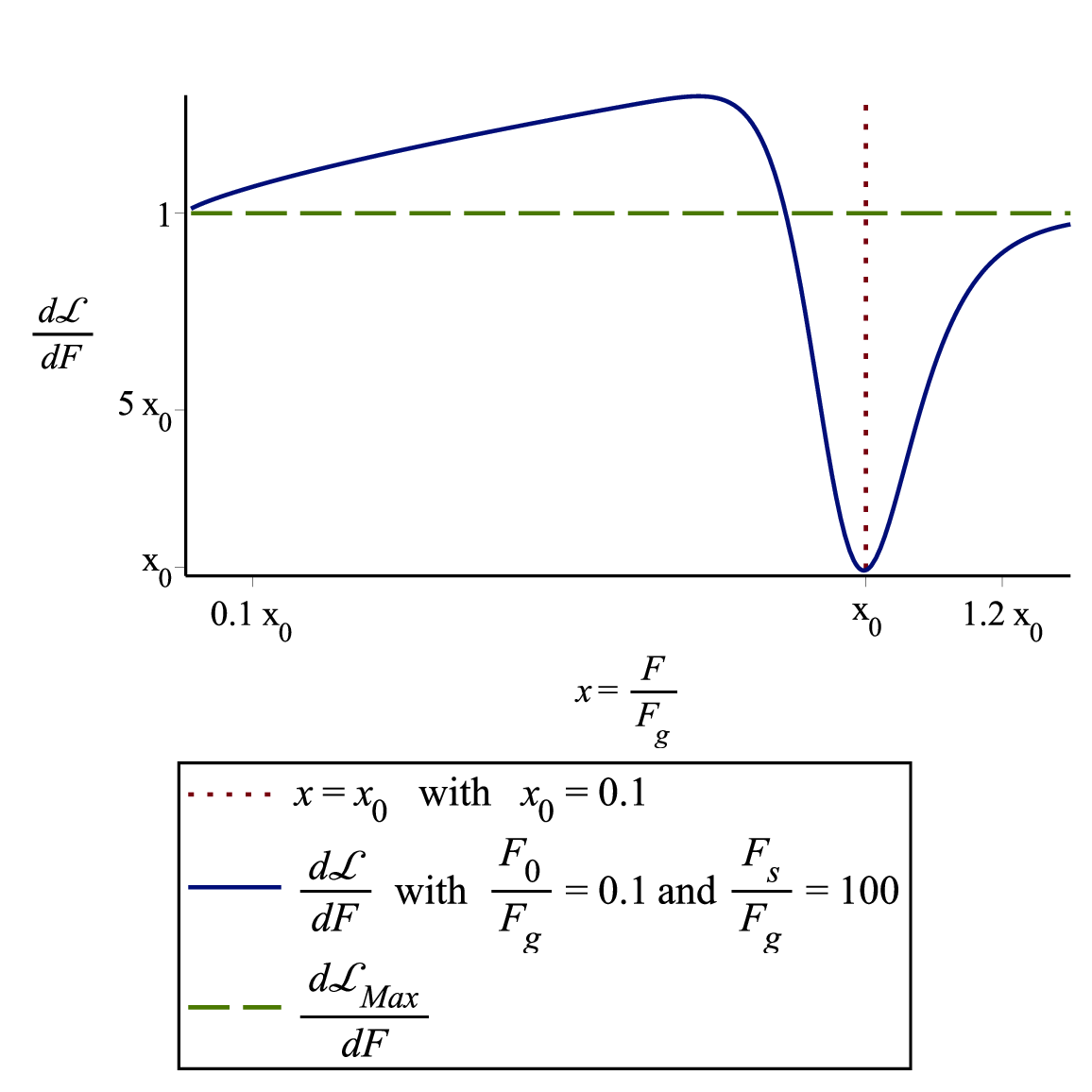}
    \caption{\label{FIG3} Behavior of $\frac{d\mathcal{L}}{d\mathcal{F}}$ 
    with respect to $x=\frac{\mathcal{F}}{\mathcal{F}_{_{g}}}\in\mathbb{R}^{+}+\{0\}$. Solid line (\textcolor{blue}{$\rule[0.8mm]{6mm}{0.3mm}$}) represents the NLED Lagrangian \eqref{NLEDmodel} with $\mathcal{F}_{_{c}}=\mathcal{F}_{_{0}}$, $x_{_{0}}=\frac{\mathcal{F}_{_{0}}}{\mathcal{F}_{_{g}}} = 10^{-1}$, $x_{_{s}}=\frac{\mathcal{F}_{_{s}}}{\mathcal{F}_{_{g}}} = 10^{2}$, and the plot basically remains the same regardless of the $v=\textup{constant}\in[10^{-4}\!~,~\!10^{-3}]$.
    Dashed line (\textcolor{OliveGreen}{$\rule[0.8mm]{3.0mm}{0.3mm}~\rule[0.8mm]{3.0mm}{0.3mm}$}) represents the Maxwell Lagrangian. Left panel shows plots in the interval $x=\frac{\mathcal{F}}{\mathcal{F}_{_{g}}}\in[0,1.3~\!x_{_{g}})$. Right panel shows plots in the interval $x=\frac{\mathcal{F}}{\mathcal{F}_{_{g}}}\in(0.01~\!x_{_{0}},1.3~\!x_{_{0}})$.
    These plots indicate that, for this setting of NLED parameters, the quantity $\mathcal{L}_{_{\mathcal{F}}}(\mathcal{F})$ determined by the NLED Lagrangian \eqref{NLEDmodel}, is positive over whole interval $\mathcal{F}\in[0,\infty)$.}
\end{figure}

\begin{figure}[H]
    \centering
    \includegraphics[width=0.35\textwidth]{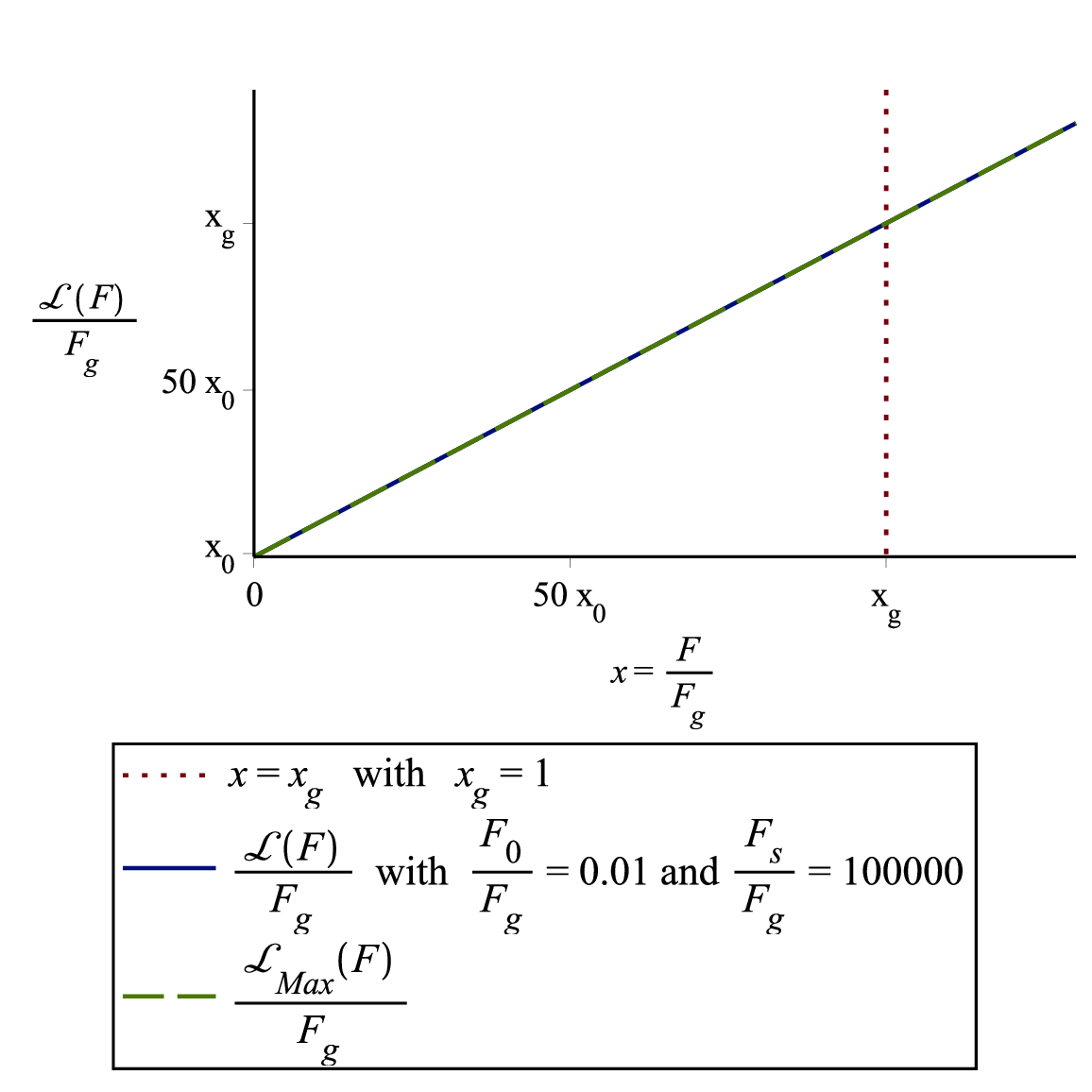}
    \hfill 
    \includegraphics[width=0.35\textwidth]{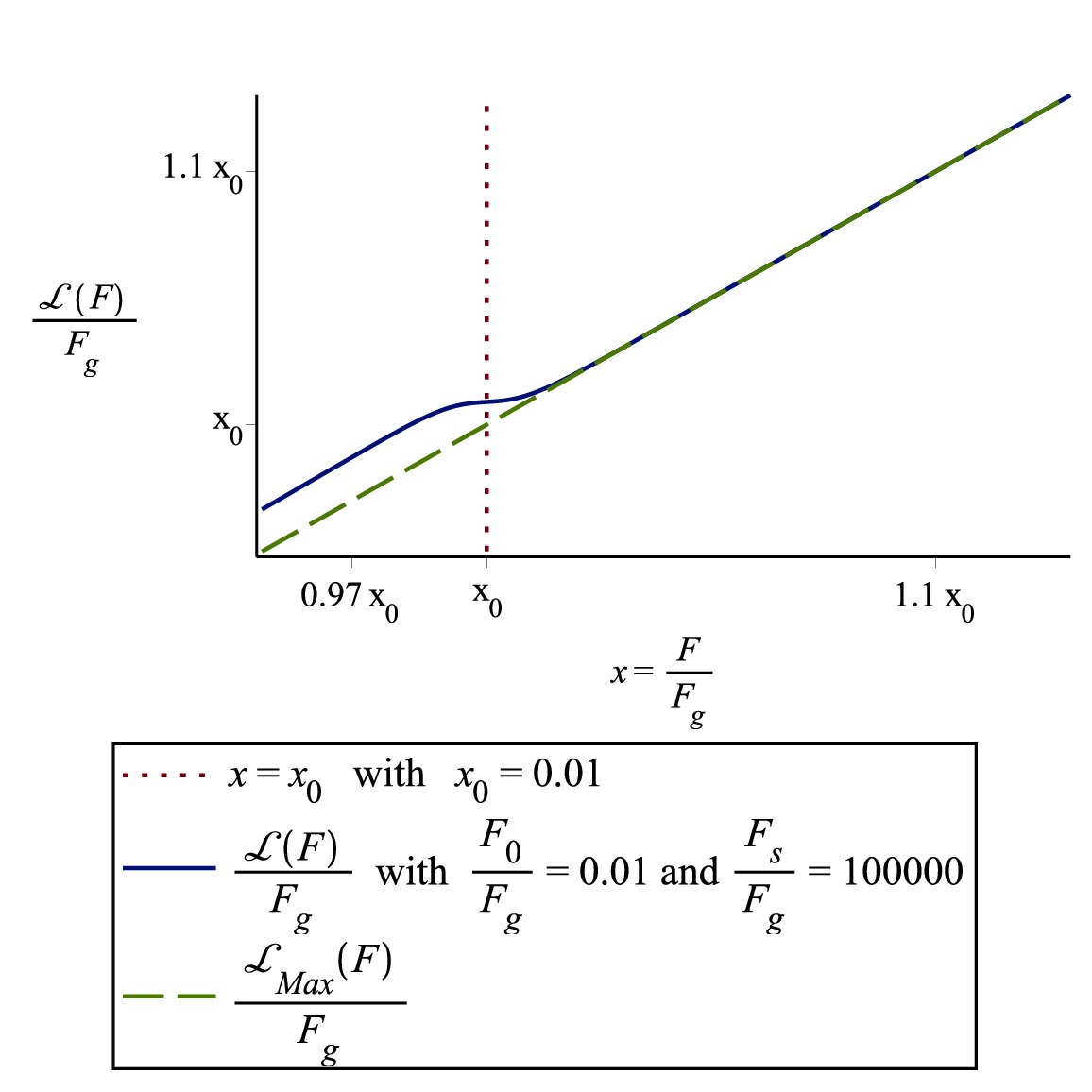}
    \caption{\label{FIG4}  The dimensionless electromagnetic Lagrangian $\frac{\mathcal{L}(\mathcal{F})}{\mathcal{F}_{_{g}}}$ is shown versus the dimensionless electromagnetic invariant $x=\frac{\mathcal{F}}{\mathcal{F}_{_{g}}}$. 
    Solid line (\textcolor{blue}{$\rule[0.8mm]{6mm}{0.3mm}$}) represents the 
    dimensionless Lagrangian for the NLED model \eqref{NLEDmodel} with $\mathcal{F}_{_{c}}=\mathcal{F}_{_{0}}$, $x_{_{0}}=\frac{\mathcal{F}_{_{0}}}{\mathcal{F}_{_{g}}} = 10^{-2}$, $x_{_{s}} = \frac{\mathcal{F}_{_{s}}}{\mathcal{F}_{_{g}}} = 10^{5}$, and the plot basically remains the same regardless of the $v=\textup{constant}\in[10^{-4}\!~,~\!10^{-3}]$. For this NLED parameter configuration, the parameters that define the solution \eqref{RCG_D} are such that $R^{4}_{_{0}}= 10^{2}~\!r^{4}_{_{g}}$, $r^{4}_{_{g}}= 10^{5}~\!(2M)^{4}$ and $v = \textup{constant} \in[10^{-4}\!~,~\!10^{-3}]$,  fulfilling  \eqref{CNDrSG}.
    Dashed line (\textcolor{OliveGreen}{$\rule[0.8mm]{3.0mm}{0.3mm}~\rule[0.8mm]{3.0mm}{0.3mm}$}) represents the
    dimensionless Lagrangian for the Maxwell electrodynamics theory. Left panel shows plots in the interval $x=\frac{\mathcal{F}}{\mathcal{F}_{_{g}}}\in[0,1.3~\!x_{_{g}})$, dot line (\textcolor{Brown}{$\cdot\!~\!\cdot\!~\!\cdot\!~\!\cdot\!~\!\cdot$}) represents the region $x = x_{_{g}} =\frac{\mathcal{F}_{_{g}}}{\mathcal{F}_{_{g}}}= 1$. Right panel shows plots in the interval $x=\frac{\mathcal{F}}{\mathcal{F}_{_{g}}}\in(0.95~\!x_{_{0}},1.13~\!x_{_{0}})$, dot line (\textcolor{Brown}{$\cdot\!~\!\cdot\!~\!\cdot\!~\!\cdot\!~\!\cdot$}) represents the region $x = x_{_{0}} =\frac{\mathcal{F}_{_{0}}}{\mathcal{F}_{_{g}}} = 10^{-2}$.
    These plots indicate that the NLED Lagrangian \eqref{NLEDmodel}, for this setting of NLED parameters, 
    is monotonically increasing and positive in the domain $\mathcal{F}\in[0,\infty)$.
    Furthermore, in this instance, it can be noted that over the whole domain $\mathcal{F}\in[0,\infty)$, with the exception of a small region around $\mathcal{F}\approx\mathcal{F}_{_{0}}$, the NLED Lagrangian \eqref{NLEDmodel} is very similar to the Maxwell Lagrangian. }
\end{figure}

\begin{figure}[H]
    \centering
    \includegraphics[width=0.35\textwidth]{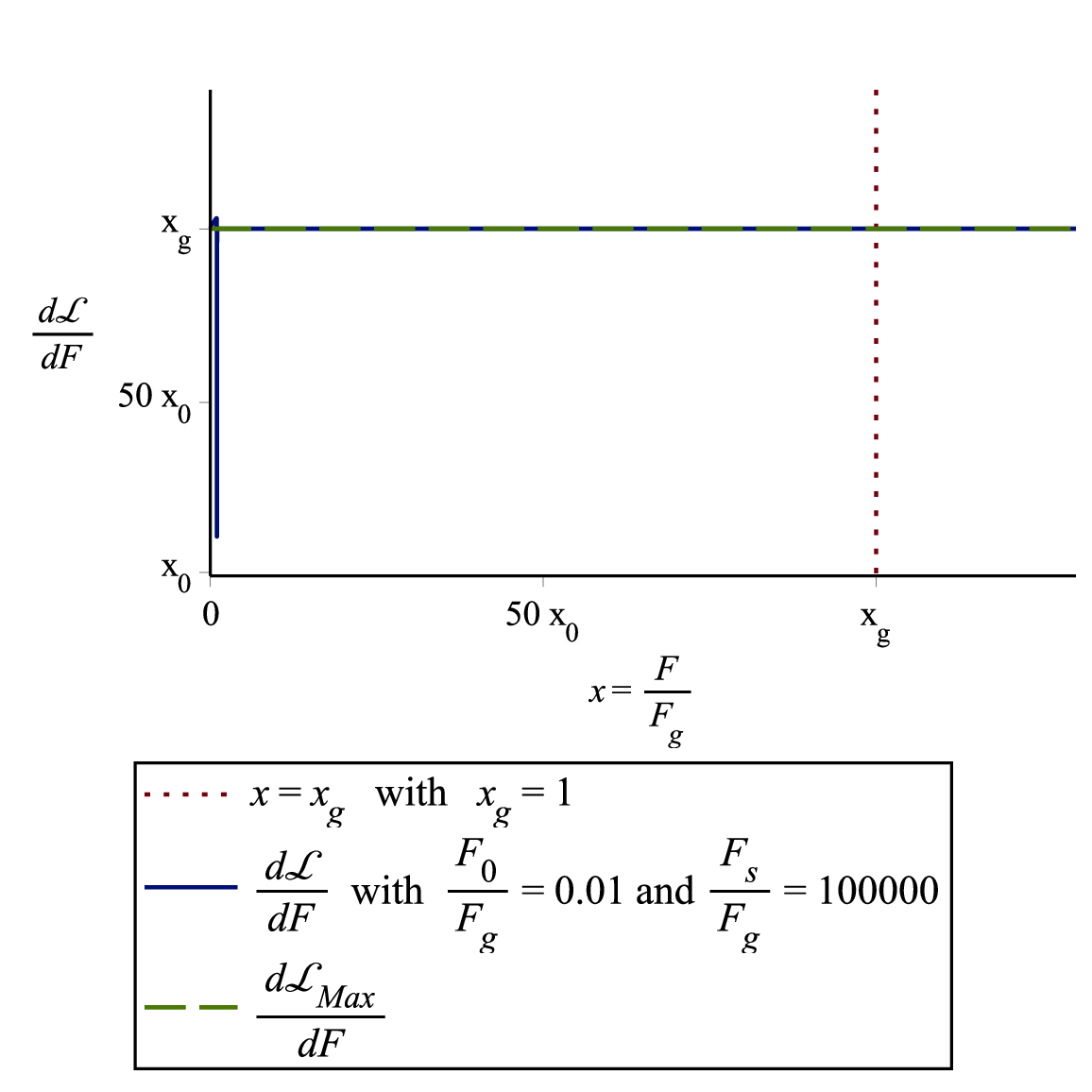}
    \hfill %
    \includegraphics[width=0.35\textwidth]{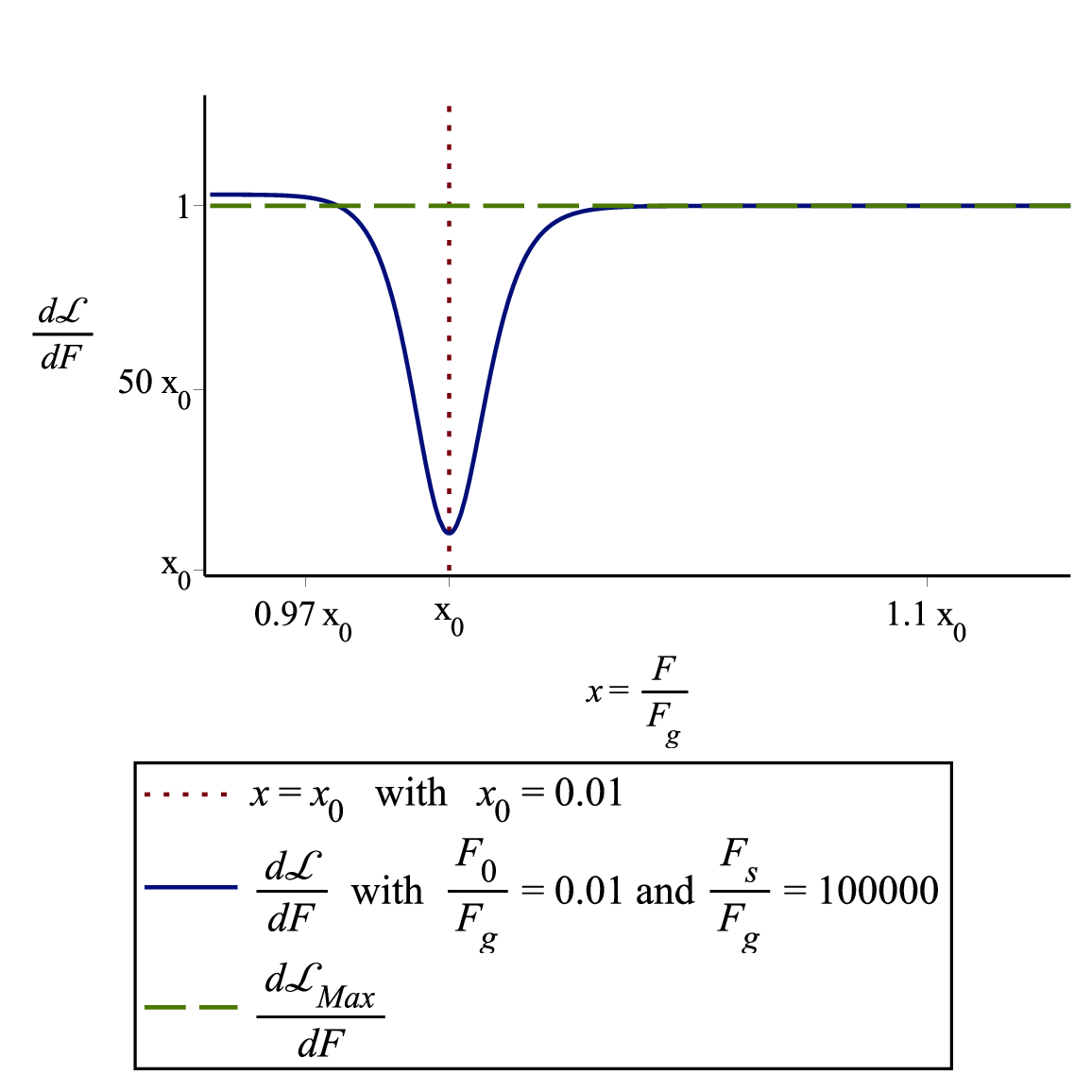}
    \caption{\label{FIG5} Behavior of $\frac{d\mathcal{L}}{d\mathcal{F}}$ with respect to $x=\frac{\mathcal{F}}{\mathcal{F}_{_{g}}}\in\mathbb{R}^{+}+\{0\}$. Solid line (\textcolor{blue}{$\rule[0.8mm]{6mm}{0.3mm}$}) represents the NLED Lagrangian \eqref{NLEDmodel} with $\mathcal{F}_{_{c}}=\mathcal{F}_{_{0}}$, $x_{_{0}}=\frac{\mathcal{F}_{_{0}}}{\mathcal{F}_{_{g}}} = 10^{-2}$, $x_{_{s}}=\frac{\mathcal{F}_{_{s}}}{\mathcal{F}_{_{g}}} = 10^{5}$, and the plot basically remains the same regardless of the 
    $v=\textup{constant}\in[10^{-4}\!~,~\!10^{-3}]$. Dashed line (\textcolor{OliveGreen}{$\rule[0.8mm]{3.0mm}{0.3mm}~\rule[0.8mm]{3.0mm}{0.3mm}$}) represents the Maxwell  
    Lagrangian. Left panel shows plots in the interval $x=\frac{\mathcal{F}}{\mathcal{F}_{_{g}}}\in[0,1.3~\!x_{_{g}})$. Right panel shows plots in the interval $x=\frac{\mathcal{F}}{\mathcal{F}_{_{g}}}\in(0.95~\!x_{_{0}},1.13~\!x_{_{0}})$. 
    These plots indicate that, for this setting of NLED parameters, the quantity $\mathcal{L}_{_{\mathcal{F}}}(\mathcal{F})$ determined by the NLED Lagrangian \eqref{NLEDmodel}, is positive over whole interval $\mathcal{F}\in[0,\infty)$.}
\end{figure}

\begin{figure}[H]%
    \centering
    \includegraphics[width=0.35\textwidth]{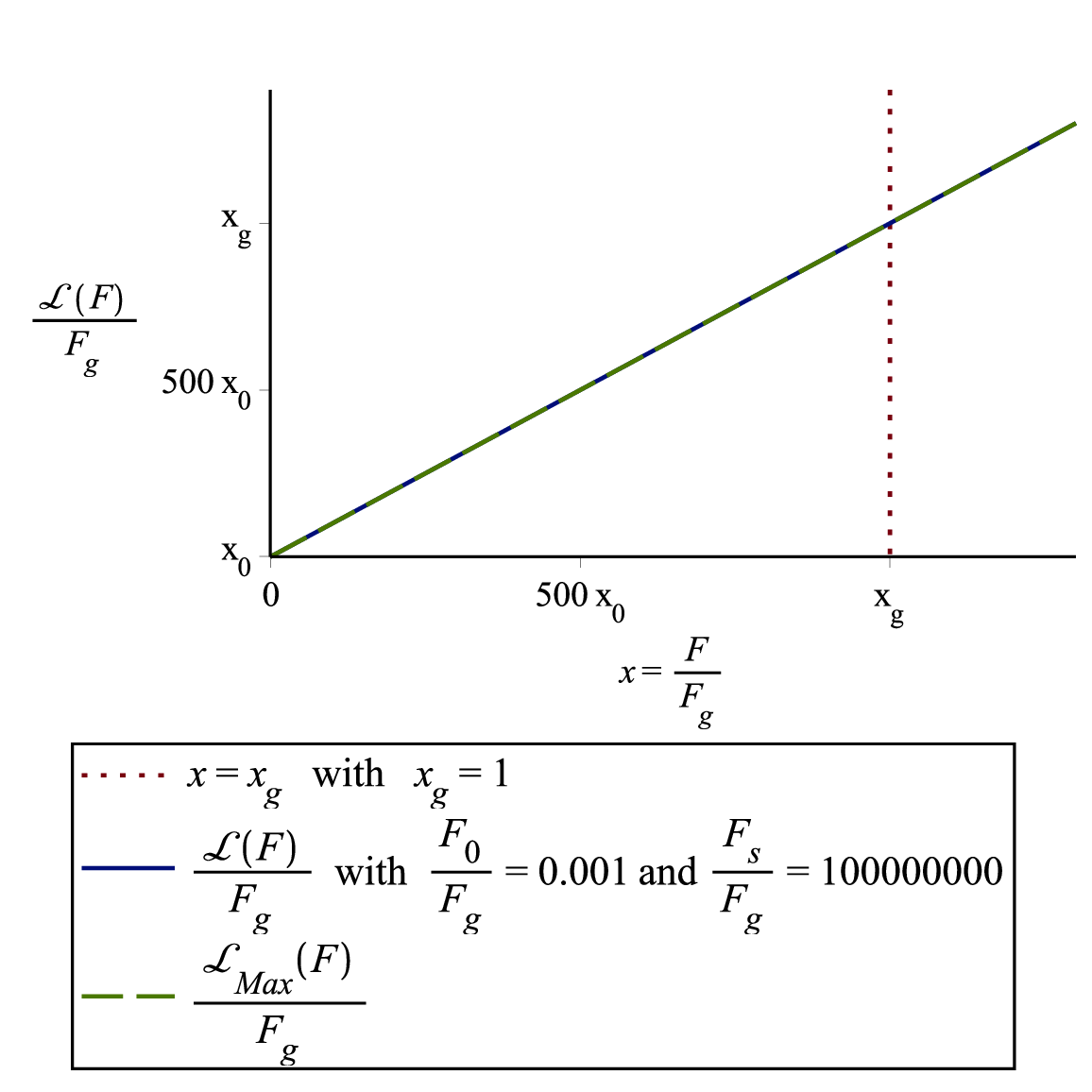}
    \hfill %
    \includegraphics[width=0.35\textwidth]{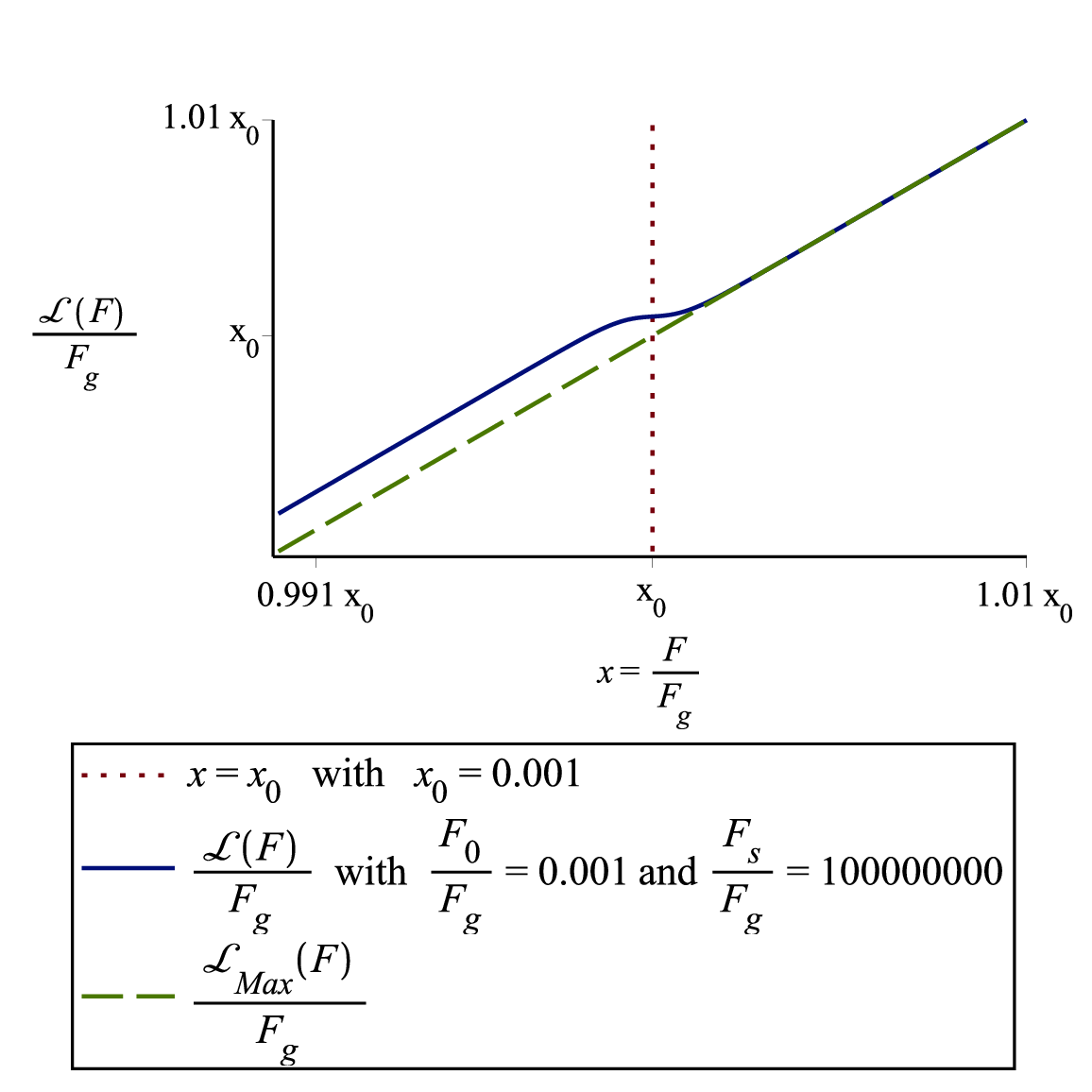}
    \caption{\label{FIG6} The dimensionless electromagnetic Lagrangian $\frac{\mathcal{L}(\mathcal{F})}{\mathcal{F}_{_{g}}}$ is shown versus the dimensionless electromagnetic invariant $x=\frac{\mathcal{F}}{\mathcal{F}_{_{g}}}$. Solid line (\textcolor{blue}{$\rule[0.8mm]{6mm}{0.3mm}$}) represents the dimensionless Lagrangian for the NLED model \eqref{NLEDmodel} with $\mathcal{F}_{_{c}}=\mathcal{F}_{_{0}}$, $x_{_{0}}=\frac{\mathcal{F}_{_{0}}}{\mathcal{F}_{_{g}}} = 10^{-3}$, $x_{_{s}} = \frac{\mathcal{F}_{_{s}}}{\mathcal{F}_{_{g}}} = 10^{8}$, and the plot basically remains the same regardless of the $v=\textup{constant}\in[10^{-4}\!~,~\!10^{-3}]$. For this NLED parameter configuration, the parameters that define the solution \eqref{RCG_D} are such that $R^{4}_{_{0}}= 10^{3}~\!r^{4}_{_{g}}$, $r^{4}_{_{g}}= 10^{8}~\!(2M)^{4}$ and $v = \textup{constant} \in[10^{-4}\!~,~\!10^{-3}]$,  fulfilling  \eqref{CNDrSG}.
    Dashed line (\textcolor{OliveGreen}{$\rule[0.8mm]{3.0mm}{0.3mm}~\rule[0.8mm]{3.0mm}{0.3mm}$}) represents the
    dimensionless Lagrangian for the Maxwell electrodynamics theory. Left panel shows plots in the interval $x=\frac{\mathcal{F}}{\mathcal{F}_{_{g}}}\in[0,1.3~\!x_{_{g}})$, dot line (\textcolor{Brown}{$\cdot\!~\!\cdot\!~\!\cdot\!~\!\cdot\!~\!\cdot$}) represents the region $x = x_{_{g}} =\frac{\mathcal{F}_{_{g}}}{\mathcal{F}_{_{g}}}= 1$. Right panel shows plots in the interval $x=\frac{\mathcal{F}}{\mathcal{F}_{_{g}}}\in(0.99~\!x_{_{0}},1.01~\!x_{_{0}})$, dot line (\textcolor{Brown}{$\cdot\!~\!\cdot\!~\!\cdot\!~\!\cdot\!~\!\cdot$}) represents the region $x = x_{_{0}} =\frac{\mathcal{F}_{_{0}}}{\mathcal{F}_{_{g}}} = 10^{-3}$.
    These plots indicate that the NLED Lagrangian \eqref{NLEDmodel}, for this setting of NLED parameters, 
    is monotonically increasing and positive in the domain $\mathcal{F}\in[0,\infty)$.
    Furthermore, in this instance, it can be noted that over the whole domain $\mathcal{F}\in[0,\infty)$, with the exception of a small region around $\mathcal{F}\approx\mathcal{F}_{_{0}}$, the NLED Lagrangian \eqref{NLEDmodel} is very similar to the Maxwell Lagrangian.}
\end{figure}

\begin{figure}[H]
    \centering
    \includegraphics[width=0.35\textwidth]{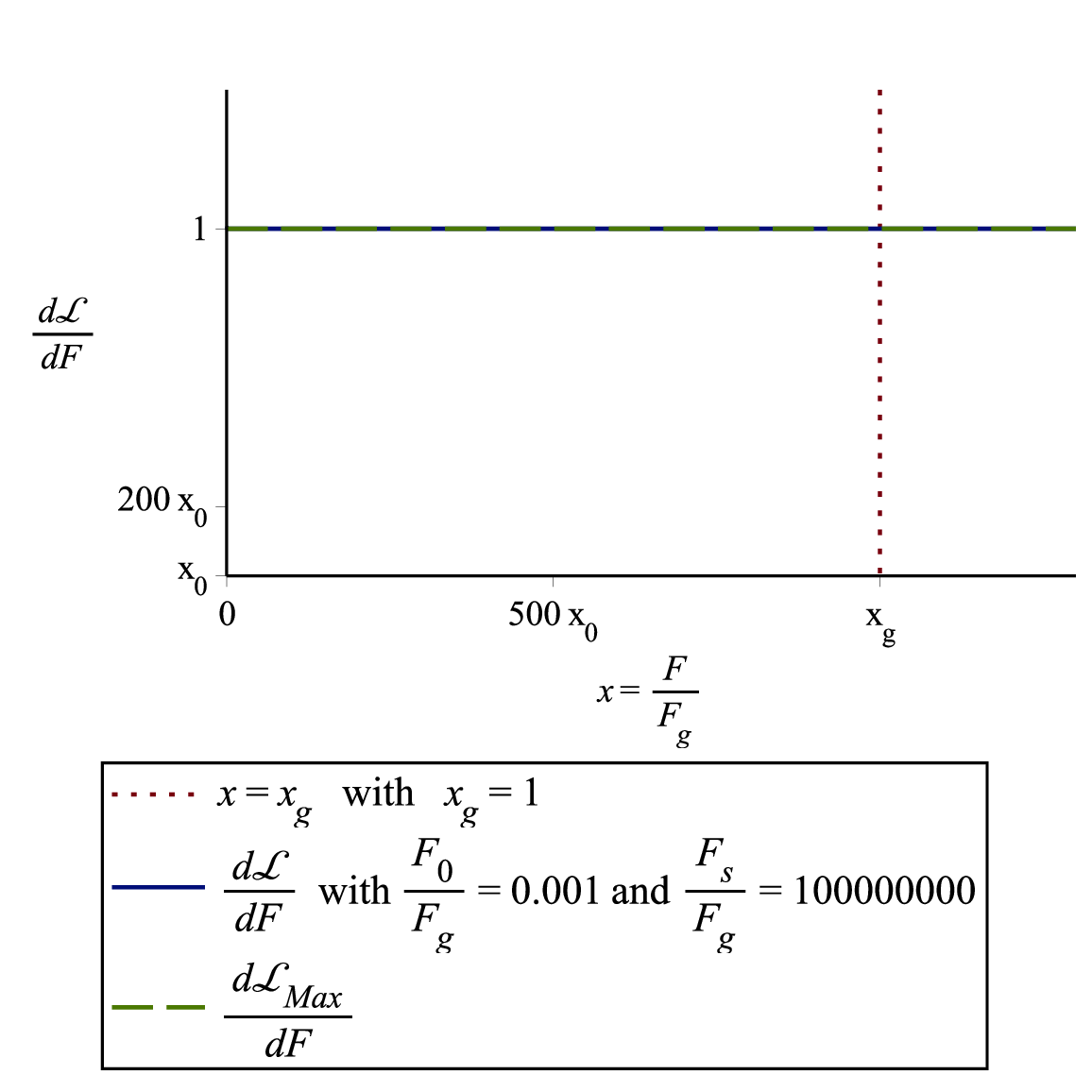}
    \hfill 
    \includegraphics[width=0.35\textwidth]{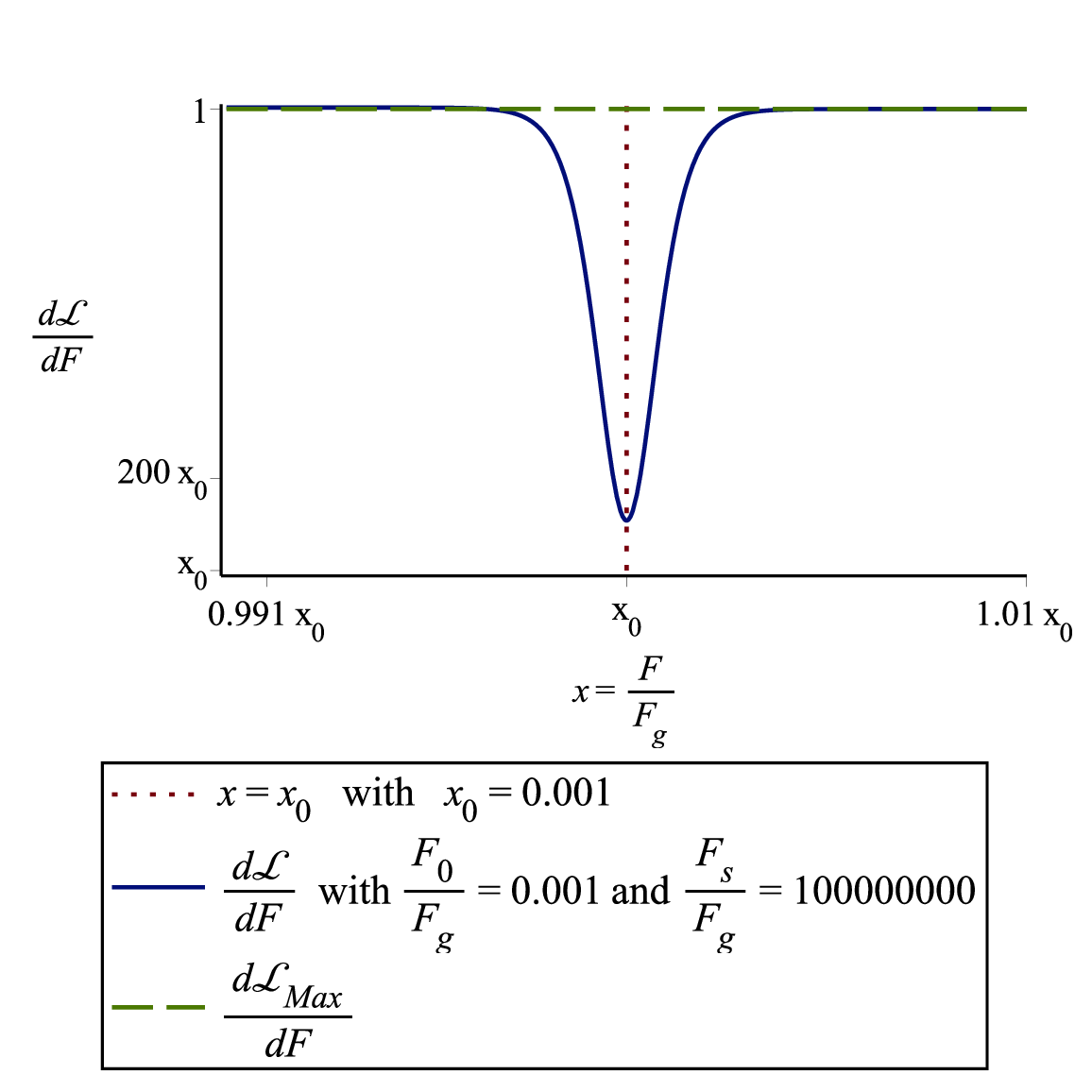}
    \caption{\label{FIG7} Behavior of $\frac{d\mathcal{L}}{d\mathcal{F}}$  with respect to $x=\frac{\mathcal{F}}{\mathcal{F}_{_{g}}}\in\mathbb{R}^{+}+\{0\}$. Solid line (\textcolor{blue}{$\rule[0.8mm]{6mm}{0.3mm}$}) represents the NLED Lagrangian \eqref{NLEDmodel} with $\mathcal{F}_{_{c}}=\mathcal{F}_{_{0}}$, $x_{_{0}}=\frac{\mathcal{F}_{_{0}}}{\mathcal{F}_{_{g}}} = 10^{-3}$,  $x_{_{s}}=\frac{\mathcal{F}_{_{s}}}{\mathcal{F}_{_{g}}} = 10^{8}$, and the plot basically remains the same regardless of the 
    $v=\textup{constant}\in[10^{-4}\!~,~\!10^{-3}]$.
    Dashed line (\textcolor{OliveGreen}{$\rule[0.8mm]{3.0mm}{0.3mm}~\rule[0.8mm]{3.0mm}{0.3mm}$}) represents the Maxwell Lagrangian. Left panel shows plots in the interval $x=\frac{\mathcal{F}}{\mathcal{F}_{_{g}}}\in[0,1.3~\!x_{_{g}})$. Right panel shows plots in the interval $x=\frac{\mathcal{F}}{\mathcal{F}_{_{g}}}\in(0.99~\!x_{_{0}},1.01~\!x_{_{0}})$. 
    These plots indicate that, for this setting of NLED parameters, the quantity $\mathcal{L}_{_{\mathcal{F}}}(\mathcal{F})$ determined by the NLED Lagrangian \eqref{NLEDmodel}, is positive over whole interval $\mathcal{F}\in[0,\infty)$.}
\end{figure}

Let us consider a photon approaching a spiral galaxy from a very great distance, $r\sim\infty$. Assuming that the gravitational field generated by the spiral galaxy is described by a metric of the form $ds^{2} =  - e^{A(r)}dt^{2} + dr^{2}  + r^{2}d\Omega^{2}$, the bending of the light can be calculated using the standard
integral-angle formula \cite{Weinberg} with $g_{_{rr}}=1$, which in this case is expressed in terms of the distance of closest approach of the photon to the center of the galaxy $r_{0}$, that is, 
\begin{equation}\label{Deltavph}
\Delta\varphi =2|\varphi(r_{0}) - \varphi_{\infty}| - \pi \quad \textup{ being } \quad \varphi(r_{0}) - \varphi_{\infty} = \int_{r_{0}}^{\infty}\left[\left(\frac{r}{r_{0}}\right)^{2}e^{A(r_{0})-A(r)} - 1 \right]^{-\frac{1}{2}}\frac{dr}{r}.   
\end{equation}
If we used the value of $A(r)$ given by \eqref{RCG}, for the case $\frac{r_{g}}{R_{0}}\sim0$, yields 
\begin{equation}\label{delta_phi}
\varphi(r_{0}) - \varphi_{\infty} \approx \int_{r_{0}}^{R_{0}} \left[\left(\frac{r}{r_{0}}\right)^{2(1-v^{2})} - 1 \right]^{-\frac{1}{2}}\frac{dr}{r} + \int_{R_{0}}^{\infty}\left[ \frac{(r_{0}-\mu)r^{3} - (r -\mu)r^{3}_{0}}{r^{3}_{0} r}\right]^{-\frac{1}{2}}\frac{dr}{r}.    
\end{equation}
using the relations
\begin{equation} 
\frac{1}{v^{2}-1}\!\!\left\{\!\tan^{-1}\!\left[\!\left(\frac{r}{r_{0}}\right)^{\!2-2v^{2}} \!- \!1 \right]^{\!-\frac{1}{2}}\!\right\}' \!=\! \frac{1}{r}\!\left[\!\left(\frac{r}{r_{0}}\right)^{\!2-2v^{2}} \!-\! 1 \!\right]^{\!-\frac{1}{2}}
\end{equation}
and
\begin{equation}
\left[\!\frac{(r_{0}\!-\!\mu)r^{3} \!-\!(r\!-\!\mu)r^{3}_{0}}{r^{3}_{0} r}\!\right]^{\!-\frac{1}{2}} \!=\!\frac{ \left[ 1 \!-\!\frac{\mu}{r}\!-\!\frac{\mu r^{2}}{r_{0}\left(r +r_{0}\right)\left(r-\mu\right)}\!+\!\frac{\mu^{2}r}{r_{0}\left(r +r_{0}\right)\left(r-\mu \right)}\right]^{\!-\frac{1}{2}} }{ \left[ \left(\frac{r}{r_{0}}\right)^{2}\!-\!1\right]^{\frac{1}{2}}}\!=\!\frac{1\!+\!\frac{\mu}{2r}\!+\!\frac{\mu r}{2r_{0}(r+r_{0}) }\!+\!\frac{3\mu^{2}(r^{2} + r_{0}r \!+\! r_{0}^{2})^{2}}{8(r+r_{0})^{2}r_{0}^{2}r^{2}}\! +\!\mathcal{O}(\mu^{3}) }{ \left[ \left(\frac{r}{r_{0}}\right)^{2} -1  \right]^{\frac{1}{2}}}
\end{equation}
to calculate\footnote{Also, we use $\left[\sin^{-1}\!\left(\!\frac{r_{0}}{r}\!\right)\right]' \!=\! - \frac{1}{r}\!\left[\!\left(\!\frac{r}{r_{0}}\!\right)^{2} \!-\!1\right]^{\!-\frac{1}{2}}$.}  the integrals \eqref{delta_phi}, we obtain
\begin{eqnarray}\label{phi_r}
\varphi(r_{0}) \!-\!\varphi_{_{\infty}} \!&\approx&\! \frac{1}{v^{2}\!-\!1}\!\!\!\left[\!\tan^{-1}\!\!\left(\!\!\frac{1}{\sqrt{\!\!\left(\!\frac{R_{0}}{r_{0}}\!\!\right)^{\!\!2-2v^{2}}\!\!-\!1~}}\!\right)\!\!-\!\!\frac{\pi}{2}\!\right]\!\!+\!\sin^{-1}\!\!\left(\!\frac{r_{0}}{R_{0}}\!\right)\! +\! \left[1\!-\!\frac{\left(2R_{0}\!+\!r_{0}\right)}{2R_{0}}\frac{\sqrt{R_{0}\!-\!r_{0}}}{\sqrt{R_{0}\!+\!r_{0}}} \right]\frac{\mu}{r_{0}}  \nonumber \\ && - \left[ \frac{1}{2} - \frac{ \left(8R_{0}^{3}+ 7 r_{0} R_{0}^{2} - 6 R_{0} r_{0}^{2} - 3r_{0}^{3} \right) \sqrt{R_{0}^{2} - r_{0}^{2}} }{16R_{0}^{2}\left(R_{0}+r_{0}\right)^{2}} - \frac{15}{16} \tan^{-1}\!\!\left(\!\!\frac{1}{\sqrt{\!\!\left(\!\frac{R_{0}}{r_{0}}\!\!\right)^{\!\!2}\!\!-\!1~}}\!\right) \right]\frac{\mu^{2}}{r_{0}^{2}} + \mathcal{O}\!\left(\frac{\mu^{3}}{r_{0}^{3}}\right).   
\end{eqnarray}
If we put $r_{0}=R_{0}$ in Eq. \eqref{phi_r}, we obtain the standard result for the Schwarzschild metric with mass $M=\mu/2$ and with $\Delta\varphi$ of Eq. \eqref{Deltavph}, up to second order in $M/R_{0}$, given by 
\begin{equation}
\Delta\varphi \approx \frac{4M}{R_{0}} + \frac{15\pi- 16}{4}\left(\frac{M}{R_{0}}\right)^{2} + \mathcal{O}\!\left(\frac{M^{3}}{R_{0}^{3}}\right).    
\end{equation}
Therefore, the final form of the spacetime metric outside the baryonic mass distribution of a spiral galaxy exhibiting flat rotation curves is obtained by substituting $r_{0}=R_{0}$ and $\mu=2M$ in equation \eqref{RCG}. The resulting metric is as follows:
\begin{equation}\label{RCG_D}
ds^{2} =  - \left[ \frac{(R_{0}-2M)r^{3} }{~(R_{0}-2M)r^{3} + 2M R^{3}_{0}~}\beta(r) + \left(\frac{r}{r_{g}}\right)^{\!\!2v^{2}}\sigma(r) \right]dt^{2} + dr^{2}  + r^{2}d\Omega^{2}, \quad\textup{ for }\quad  r \in (0,\infty)
\end{equation}
with $\sigma(r)$ given by Eq. \eqref{sigma1r} and $\beta(r)=1-\sigma(r)$. \\
Finally, let us emphasize that the purely magnetic spacetime metric 
\eqref{RCG_D} with parameters such that 
\begin{equation}\label{CNDrSG}
R^{4}_{_{0}} \gg r^{4}_{_{g}} \gg (2M)^{4} 
\quad\textup{ and }\quad v\in[10^{-4}\!~,~\!10^{-3}]  
\end{equation}
which describes the spacetime geometry of a spiral galaxy, comes from considering an E-STGB-NLED theory with NLED Lagrangian density given by Eq. \eqref{NLEDmodel}, defined on $\mathcal{F}\in[0,\infty)$, and with parameters such that: 
\begin{equation}\label{NLEDparameters}
\mathcal{F}_{_{c}} =\mathcal{F}_{_{0}} = \frac{q^{2}}{2R^{4}_{_{0}}}, \quad \mathcal{F}_{_{s}} = \frac{q^{2}}{2(2M)^{4}}, \quad \mathcal{F}_{_{g}} = \frac{q^{2}}{2r^{4}_{_{g}}}, \quad \mathcal{F}_{_{s}} \gg \mathcal{F}_{_{g}} \gg  
\mathcal{F}_{_{0}} > 0, \quad\textup{ and }\quad v\in[10^{-4}\!~,~\!10^{-3}].    
\end{equation}
For this setting of parameters, the NLED Lagrangian density \eqref{NLEDmodel} is positive-valued in the whole domain of the solution \eqref{RCG_D}.
Moreover, in this instance, the NLED model \eqref{NLEDmodel} closely approximates Maxwell electrodynamics ({\it i.e.,} $\mathcal{L}(\mathcal{F})\approx\mathcal{F}$) in the domain $\mathcal{F}\geq0$ with the exception of a small region around $\mathcal{F}=\mathcal{F}_{_{0}}$ where this NLED model interpolates between \eqref{out_NLED} and \eqref{ins_NLED}, see Figs.~\!\ref{FIG2}-\ref{FIG7}. 
More specifically, provided the conditions \eqref{NLEDparameters} are fulfilled, 
the function $\sigma(\mathcal{F})$ exhibits the behavior described in Eq. \eqref{sigma1Bh}. Consequently, the NLED model \eqref{NLEDmodel} will interpolates between two NLED models: the model \eqref{out_NLED} defined on $\mathcal{F}\in[0,\mathcal{F}_{_{0}}-\epsilon^{2})$, which represents $\mathcal{F}$ values outside the spiral galaxy, and the model \eqref{ins_NLED} defined on $\mathcal{F}\in(\mathcal{F}_{_{0}}+\epsilon^{2},\infty)$, which represents $\mathcal{F}$ values inside the FRC zone.
The dimensionless Lagrangian for the NLED model \eqref{NLEDmodel} with $\mathcal{F}_{_{c}}=\mathcal{F}_{_{0}}$, is given by
\begin{equation}\label{NLEDmodelSGx}
\frac{\mathcal{L}(\mathcal{F})}{\mathcal{F}_{_{g}}} = \left\{ \frac{\left[\left(\frac{\mathcal{F}_{_{0}}}{\mathcal{F}_{_{g}}}\right)^{-\frac{1}{4}} - \left(\frac{\mathcal{F}_{_{s}}}{\mathcal{F}_{_{g}}}\right)^{-\frac{1}{4}} \right]\left(\frac{\mathcal{F}}{\mathcal{F}_{_{g}}} \right)^{-\frac{3}{4}} }{~\left[\left(\frac{\mathcal{F}_{_{0}}}{\mathcal{F}_{_{g}}}\right)^{-\frac{1}{4}} - \left(\frac{\mathcal{F}_{_{s}}}{\mathcal{F}_{_{g}}}\right)^{-\frac{1}{4}}\right] \left(\frac{\mathcal{F}}{\mathcal{F}_{_{g}}}\right)^{-\frac{3}{4}} + \left(\frac{\mathcal{F}_{_{s}}}{\mathcal{F}_{_{g}}}\right)^{-\frac{1}{4}} \left(\frac{\mathcal{F}_{_{0}}}{\mathcal{F}_{_{g}}}\right)^{-\frac{3}{4}}~}\beta(\mathcal{F}) + \left(\frac{\mathcal{F}_{_{g}}}{\mathcal{F}}\right)^{\!\!\frac{v^{2}}{2}}\!\sigma(\mathcal{F}) \right\}^{\!\!\!^{-1}}\frac{\mathcal{F}}{\mathcal{F}_{_{g}}} 
\end{equation}
where $\beta(\mathcal{F}) = 1 - \sigma(\mathcal{F})$ and $\sigma(\mathcal{F}) = \frac{1}{2}  - \frac{1}{2} \tanh\!\left(\frac{\mathcal{F}_{_{g}}}{\mathcal{F}} - \frac{\mathcal{F}_{_{g}}}{\mathcal{F}_{_{0}}}\right)$. 
The figures above show that the NLED model \eqref{NLEDmodel} defined for  
$\mathcal{F}\in[0,\infty)$, with $v = \textup{constant} \in[10^{-4}\!~,~\!10^{-3}]$, $\mathcal{F}_{_{c}}=\mathcal{F}_{_{0}}$ and the following selection of relative parameters $\left\{\frac{\mathcal{F}_{_{0}}}{\mathcal{F}_{_{g}}}, \frac{\mathcal{F}_{_{s}}}{\mathcal{F}_{_{g}}} \right\}$  that satisfy \eqref{CND}, given by;  $\frac{\mathcal{F}_{_{0}}}{\mathcal{F}_{_{g}}} = 10^{-1}$  and  $\frac{\mathcal{F}_{_{s}}}{\mathcal{F}_{_{g}}} = 10^{2}$, used in Figs.~\!\ref{FIG2}-\ref{FIG3}, NLED source of spiral galaxy spacetime \eqref{RCG_D} with $\frac{R^{4}_{_{0}}}{r^{4}_{_{g}}} = 10$  and  $\frac{r^{4}_{_{g}}}{(2M)^{4}} = 10^{2}$; $\frac{\mathcal{F}_{_{0}}}{\mathcal{F}_{_{g}}} = 10^{-2}$ and $\frac{\mathcal{F}_{_{s}}}{\mathcal{F}_{_{g}}} = 10^{5}$, used in Figs.~\!\ref{FIG4}-\ref{FIG5}, NLED source of spiral galaxy spacetime \eqref{RCG_D} with $\frac{R^{4}_{_{0}}}{r^{4}_{_{g}}} = 10^{2}$  and $\frac{r^{4}_{_{g}}}{(2M)^{4}} = 10^{5}$; $\frac{\mathcal{F}_{_{0}}}{\mathcal{F}_{_{g}}} = 10^{-3}$ and $\frac{\mathcal{F}_{_{s}}}{\mathcal{F}_{_{g}}} = 10^{8}$, used in Figs.~\!\ref{FIG6}-\ref{FIG7}, NLED source of spiral galaxy spacetime \eqref{RCG_D} with $\frac{R^{4}_{_{0}}}{r^{4}_{_{g}}} = 10^{3}$  and  $\frac{r^{4}_{_{g}}}{(2M)^{4}} = 10^{8}$,  satisfies the NEC and WEC by having $\mathcal{L}(\mathcal{F})\geq0$ and $\mathcal{L}_{_{\mathcal{F}}}(\mathcal{F})\geq0$. Furthermore, the figures show that for these NLED parameters, the NLED model is surprisingly similar to Maxwell electrodynamics, {\it i.e.,} $\mathcal{L}(\mathcal{F})\approx\mathcal{F}$ for almost all $\mathcal{F}\in[0,\infty)$.
\section{Discussion and Summary} 
In the present work, within the framework of E-STGB-NLED gravity, a very simple alternative to the galactic dark matter problem has been presented. Specifically, we have constructed a novel type of SSS-AF purely magnetic spacetime metric given by Eq. \eqref{RCG_D} which is capable of explaining flat rotational curves and the gravitational lensing observed in spiral galaxies.
Our explanation of the origin of the FRC zone is based on the assumption that both stars and hydrogen gas clouds, moving in large circular orbits around the galactic center, behave as massive test particles that follow timelike geodesics in the spacetime metric \eqref{RCG_D}.
This spacetime geometry characterizes the FRC zone as a thin disk-shaped region (located in the equatorial plane, $\theta=\pi/2$, centered at $r = 0$) of radius $R_{_{0}}$,
within which massive test particles traveling in circular paths around $r=0$ maintain a constant tangential velocity (which is independent of the radial distance $r$) expressed as $v = \textup{constant} \in[10^{-4}\!~,~\!10^{-3}]$ for all $r\in(0, R_{0})$. \\
The above represents a simple alternative to the galactic dark matter problem within the framework of E-STGB-NLED gravity, since the line element \eqref{RCG_D}  
can be interpreted as an exact purely magnetic solution of the  E-STGB-NLED field equations for SSS spacetimes, with NLED Lagrangian density given by \eqref{NLEDmodel}, and E-STGB theory determined by \eqref{sol_ph}$-$\eqref{sol_a}.

On the other hand, in the framework of general relativity, the metric \eqref{RCG_D} 
can be interpreted as an exact solution of the Einstein field equations, $G_{\mu}{}^{\nu} = 8\pi T_{\mu}{}^{\nu}$, with an energy-momentum tensor of the form $8\pi T_{\mu}{}^{\nu} = \textup{diag}(0,~ p_{r},~ p_{\tau},~ p_{\tau})$ representing an exotic source  
having zero rest energy density ($\rho =0$), non-zero radial and tangential pressures ($p_{r}\neq0$ and $p_{\tau}\neq0$) given, respectively, by  
\begin{equation}\label{sourceGR}
\rho = 0, ~~~ p_{r} =  \frac{A'}{r} ~~~\textup{ and }~~~ p_{\tau} = \frac{2A' + rA'^{2} + 2rA'' }{4r}.    
\end{equation}
Therefore, contrary to the dark matter hypothesis, the field source for the spiral galaxy spacetime \eqref{RCG_D} in the GR framework can be described as a ``pure pressure fluid,'' {\it i.e.,} an anisotropic fluid with zero rest energy density in which the pressures ($p_{r}$ and $p_{\tau}$) constitute the only source of energy.
This is because the standard dark matter approach accounts for the uniformity of spiral galaxy rotation curves by suggesting that the spiral galaxies are embedded in spherical halos of 
non-luminous matter (known as dark matter). 
This hypothetical matter is required to have a positive rest energy density, $\rho_{_{DM}}$, which decreasing as $1/r^{2}$, unlike the field source \eqref{sourceGR} which maintains a rest energy density equal to zero throughout the whole spiral galaxy spacetime. \\
In conclusion, the novel type of static and spherically symmetric solution \eqref{RCG_D} satisfies two crucial physical requirements: the gravitational lensing and rotation curves of spiral galaxies. 
Investigations are currently in progress to examine the constraints of this model, particularly concerning the stability of this novel type of solution.

\end{document}